\documentclass[10pt,reqno,final]{article}
\pdfoutput=1
\usepackage{amsmath,amsfonts,amssymb,amsthm,version}
\usepackage{mathrsfs,fancybox,pifont}
\usepackage{graphicx} 
\usepackage{float}     
\usepackage{url,hyperref}
\usepackage[notcite,notref]{showkeys}
\usepackage{color}
\usepackage{subfigure,multirow}
\usepackage{epstopdf}
\usepackage{cases}
\usepackage{mathtools}
\usepackage{algorithm,algorithmic}
\usepackage{authblk}
\usepackage{lipsum}
\usepackage{verbatim}
\usepackage{array}
\allowdisplaybreaks

\numberwithin{equation}{section}
\numberwithin{figure}{section}
\numberwithin{table}{section}

\theoremstyle{plain}

\theoremstyle{definition}

\theoremstyle{remark}

\newcommand{\RNum}[1]{\uppercase\expandafter{\romannumeral #1\relax}}

\title{A numerical simulation method of fish adaption behavior based on deep reinforcement learning and fluid-structure coupling -- realization of some lateral line functions\thanks{The work was supported in part by the Program of national natural science Fundation for young scholars (52109150) and by the Science and Technology Research Program of Chongqing Municipal Education Commission(Grant no. KJQN201900748).} }
\author[1,2]{Tao Li}
\author[1,2,3]{Chunze Zhang\thanks{zhangchunze@whu.edu.cn}}
\author[1,2,3]{Peiyi Peng}
\author[1,2,3]{Ji Hou}
\author[1,2,3]{Qin Zhou}
\author[1,2,3]{Qian Ma}
\affil[1]{Chongqing Southwestern Research Institute for Water Transport Engineering, Chongqing Jiaotong University}
\affil[2]{Key Laboratory of Inland Waterway Regulation Engineering of Ministry of Communications, Chongqing Jiaotong University}
\affil[3]{Chongqing Xike Consultation Center for Water Transport}

\date{January 5, 2023}

\begin{document}
	\maketitle

	\begin{abstract}
		Improving the numerical method of fish autonomous swimming behavior in complex environments is of great significance to the optimization of bionic controller, the design of fish passing facilities and the study of fish behavior. This work has built a fish autonomous swimming simulation platform, which adapts the high-precision immersed boundary–Lattice Boltzmann method (IB-LBM) to simulate the dynamic process of the interaction between the fish and the flow field in real time, and realizes the fish brain motion control through the Soft Actor-Critic (SAC) deep reinforcement learning algorithm. More importantly, in view of the poor generalization of the existing simulation platform, a method to simulate the fish's lateral line function is proposed. By adding the Lateral-line machine and designing the Macro-action system, the intelligent fish initially has the ability to recognize, classify, memorize and transplant the corresponding swimming strategy in the unsteady field. Using this method, the training and simulation of point-to-point predation swimming and K$\acute{a}$m$\acute{a}$n-gait test under different inlet velocities are carried out. In the example of point-to-point predation swimming, the fish in random position can adjust the swimming posture and speed autonomously to catch the fast moving food, and has a certain prediction ability on the movement trajectory of the food. In the K$\acute{a}$m$\acute{a}$n-gait test, the trained fish are placed in three different K$\acute{a}$m$\acute{a}$n-gait flow fields, to study its ability to recognize the flow field and select swimming strategies through experience. The results of numerical experiments show that, comparing with the other value function networks, the SAC algorithm based on maximum entropy RL framework and off-policy has more advantages in convergence speed and training efficiency when simulating fish brain decision-making. The use of the Lateral-line Machine and Macro-action system can avoid the waste of experience and improve the adaptability of intelligent fish in the new complex flow field environment.
		

	\end{abstract}

	\section{Introduction}
	\paragraph{}
	The advancement in bionics research continually presents new concepts and opportunities for the creation of cutting-edge autonomous underwater vehicles(AUV) and wave energy converters(WEC) among other underwater special equipment\cite{jalvingNDREAUVFlightControl1994,paullAUVNavigationLocalization2014,huangAUVAssistedDataGathering2020,Huang2020,Zhang2020,Wang2022}. The swimming patterns of fish in both freshwater and marine environments are renowned for their high propulsion efficiency and low energy consumption. Therefore, investigating the swimming mechanism of fish and understanding the relationship between fish and the flow field is crucial in developing innovative bionically-inspired AUVs and WECs.

	Initial researchers\cite{liaoFishExploitingVortices2003,liaoKarmanGaitNovel2003,liaoNeuromuscularControlTrout2004,liaoRoleLateralLine2006}primarily focused on investigating the swimming mechanism of fish through physical experimentation. They conducted statistical analysis on the collected experimental data, yielding valuable insights into the appearance, swimming patterns, and habits of fish. While physical experiments can establish fundamental theories of fish kinematics, kinetics, and hydrodynamics, they are often accompanied by drawbacks such as high costs and difficulties in controlling live fish. In contrast,Computational Fluid Dynamics(CFD) methods offer greater flexibility, the ability to simulate complex physical processes repeatedly, and the capability to fully control the behavior of virtual fish. As a result, CFD methods have gained widespread attention in recent years.

	Traditionally, numerical simulations using CFD have concentrated on the development of fluid-structure coupled solvers that are suitable for simulating fish swimming, with a focus on enhancing solver’s accuracy and computational speed\cite{carlingSELFPROPELLEDANGUILLIFORMSWIMMING,leroyerNumericalMethodsRANSE2005,kernSimulationsOptimizedAnguilliform2006,xinVorticityDynamicsControl2018}. Researchers have also achieved a number of academic achievements in this field. However, traditional fish swimming simulation methods can only simulate the passive motion of the fish body, with the swimming behavior being pre-determined. In order to gain a deeper understanding of the fundamental behavior of fish swimming, and to explain it from the perspective of intelligent life, it is necessary to incorporate intelligent control algorithms into traditional CFD algorithms.

	With the growing popularity of intelligent algorithms such as machine learning in recent years, the numerical simulation method of fish swimming has entered a new stage. Verma et al. \cite{vermaEfficientCollectiveSwimming} firstly introduced reinforcement learning algorithms into the numerical simulation of fish swimming, using a high-precision DNS method coupled with a value-based deep reinforcement learning algorithm to study the energy-saving mechanism of fish swimming in vortices, providing a valuable sample for the study of the fish swimming mechanism. Zhu et al. \cite{zhuNumericalStudyFish2021} investigated the swimming behavior of intelligent fish, such as point-to-point swimming, fish reorientation, and Kámán gaiting, using a LSTM-DRQN deep learning reinforcement learning algorithm coupled with the immersed boundary-lattice boltzmann method (IB-LBM). Their work incorporates fish position, velocity, and acceleration into the state space of deep recurrent Q network(DRQN), and considers the amplitude and frequency of action space as well as historical effects. Yan et al. \cite{yanNumericalSimulationMethod2020}  developed a fish swimming solver based on an overlapping lattice algorithm and coupled it with a deep reinforcement learning algorithm to achieve motion control of fish swimming. They trained a simplified two-dimensional fish navigation module, enabling the fish to follow a given path.

	The above studies have provided valuable insights for interdisciplinary research in artificial intelligence and computational fluid dynamics fields, but the application of deep reinforcement learning algorithms is still in its infancy. Specifically, simply inputting some characteristic parameters of the flow field into the state space does not enable the intelligent fish to recognize the feature of the flow field, and the swimming policy cannot be transferred across multiple flow fields. Our objective is to establish coupled modeled include lateral-line machine and marco-action system,in order to obtain a more robust intelligent fish swimming controller that can realize the agent adapt to different upstream velocities in turbulent flow fields and the knowledge transfer process of reinforcement learning.

	The structure of this paper is as follows: In the section "Numerical model and methodology," we will introduce the fluid-structure interaction algorithm and the deep reinforcement learning algorithm. In the section "Special model development," we will delve into the principle of the lateral-line machine and macro-action system in DRL in detail. In the section "Results", we will simulate predation swimming test and Kámán gating swimming tests in different flow fields to verify the reliability of the new simulator coupled with two special modules. Finally, in the section "Conclusion", we will summarize this work and outline future research directions.

	\section{Numerical model and methodology}
		\subsection{The immersed boundary–Lattice Boltzmann coupling scheme by iterative force correction}

		\paragraph{}
		The immersed boundary lattice Boltzmann method is a highly-efficient and accurate fluid-structure interaction simulation method that falls under the category of interface-capturing numerical methods. It is particularly well-suited for simulating fluid-structure interaction problems involving large deformations, such as fish swimming, insect winging, and heartbeats(\cite{tongCoarsegrainedAreadifferenceelasticityMembrane2018,suzukiEffectWingMass2019,qingInfluenceChannelRegulating2021}). It has been widely applied to a variety of fluid-structure interaction problems with great success.
		\paragraph{}
		In the immersed boundary lattice-Bolzmann method, the fluid flow is governed by the discrete lattice-Bolzmann equation \ref{equ_1}, in which the Cheng's force term\cite{chengIntroducingUnsteadyNonuniform2008} is incorporated, which results in second-order accuracy:
		\begin{equation}\label{equ_1}
			\begin{array}{l}f_a^d(x+e_a\Delta t,t+\Delta t)-f_a^d(x,t)\\ =-M^{-1}S[m_a(x,t)-m_a^{eq}(x,t)]+\dfrac{\Delta t}{2}\left[g_a^d(x,t)+g_a^d(x+e_a\Delta t,t+\Delta t)\right]\end{array}
		\end{equation}

		\paragraph{}
		In equation \ref{equ_1}, the term with superscript \textbf{d} indicates that the value is the ideal reference value, and strictly satisfies the no-slip boundary condition; \emph{t} represents the current moment of the transient model;$\Delta t$  is the given calculation time step; ${f_a}^d(x,t):\alpha  = 0,1,...,8$  denotes the particle equilibrium distribution function that strictly satisfies the no-slip boundary condition at the position and moment t; \textbf{\emph{M}} is the collision matrix; $\hat{S}$ is the diagonal matrix, corresponding to the relaxation factor in the MRT model;${m_a}(x,t):\alpha  = 0,1,...,8$   is the equilibrium distribution function of the flow field lattice point at the position of moment \textbf{t}. The relationship between ${m_a}(x,t)$  and ${f_a}^d(x,t)$  can be expressed as ${m_a}(x,t) = M{f_a}^d(x,t)$ ;${m_a}^{eq}(x,t)$  is the  ${m_a}(x,t)$’s equilibrium state in the moment space;${g_a}^d(x,t):\alpha  = 0,1,...,8$ is the external force on the fluid point at the position \textbf{x} and moment \emph{t}.
		
		\paragraph{}
		For the fluid-interface interaction process in the immersed boundary method:
		
		\begin{equation}\label{equ_2}
			{\bf{U}}(s,t) = \int_{{\Omega _{\rm{f}}}} {\bf{u}} (x,t)\delta (x - {\bf{X}}(s,t)){\rm{d}}{\bf{x}}
		\end{equation}	
	
		\begin{equation}\label{equ_3}
			{\bf{f}}(x,t) = \int_{{\Gamma _{\rm{b}}}} {\bf{F}} (s,t)\delta (x - {\bf{X}}(s,t)){\rm{d}}s
		\end{equation}
	
		\paragraph{}
		Where:lowercase letters denote the Eulerian variables defined on the fluid region $\Omega_{f}$, uppercase letters denote the Lagrangian variables defined on the immersed boundary $\Gamma_{b}$,${\bf{x}}$ , ${\bf{f}}$ and   ${\bf{u}}$ are the position vector in Cartesian coordinates, the external force term in the N-S equation, and the velocity of the Eulerian variables, respectively.${\bf{X}}$ , ${\bf{F}}$ and ${\bf{U}}$  are the position of the boundary in Lagrangian coordinates, the force of the immersed boundary on the fluid, and the velocity of the interface, respectively.   $\delta$ is the Dirac delta function, which is the key function for the implementation of fluid-structure interaction, and it can interpolate the fluid velocity ${\bf{u}}$ in equation \ref{equ_2} to get the velocity of the immersed boundary ${\bf{U}}$, or diffuse the force ${\bf{F}}$ of the immersed boundary to the surrounding fluid nodes in equation \ref{equ_3} to obtain ${\bf{f}}$.
		
		\paragraph{}
		In the immersed boundary method, the accuracy of the force term ${\bf{F}}$ plays a crucial role in satisfying the no-slip boundary condition. The method proposed by Zhang et al.\cite{zhangAccuracyImprovementImmersed2016}modifies the immersed boundary force through a series of iterations to approximate the true value, which is calculated as:
		\begin{equation}\label{equ_4}
			{F^{t + 1}}(s,t) = {F^t}(s,t) + 2\frac{{{U^d}(s,t) - {U^t}(s,t)}}{{\Delta t}}
		\end{equation}
	
		\paragraph{}
		In equation \ref{equ_4}: \textit{t} represents the number of iteration steps;${U^d}(s,t)$ is the actual velocity of the immersed boundary in the fluid;  ${U^t}(s,t)$ is the temporary velocity of the t-th iteration step;${F^t}(s,t)$ is the force term of the t-th iteration step;$\Delta t$  is the time step length.
		
		\paragraph{}
		By using equation \ref{equ_4}, the value of ${\bf{F}}$ can be calculated precisely after the iteration process, and then the final iteration's value is incorporated into equation \ref{equ_1}.
		
		\paragraph{}
		For the two-dimensional immersed boundary method of fluid-structure interaction problem, it is necessary to take into account the translation and rotation of the immersed boundary. The control equation is as follows:
		\begin{equation}\label{equ_5}
			M\frac{{{\rm{d}}{{\bf{U}}_{\bf{s}}}}}{{{\rm{dt}}}}{\rm{ = }}\int_{\Gamma b} {{\bf{F(s,t)}}ds} 
		\end{equation}
	
		\begin{equation}\label{equ_6}
			I\frac{{{\rm{d}}{\bf{\omega }}}}{{dt}} = {{\bf{T}}_{\bf{s}}}
		\end{equation}
	
		\paragraph{}
		Where: ${\bf{U_{s}}}$  is the velocity of the immersed boundary motion;$\Gamma_{b}$   represents the body function of the immersed boundary; \textit{M} is the mass of the immersed boundary, $\int_{\tau b} {F(s,t)ds}$  is the combined force of the hydrodynamic force acting on the immersed boundary; ${\bf{T_{s}}}$  is the torque on the immersed boundary; \textit{I} is the rotational inertia of the immersed boundary; $\bf{\omega}$  is the rotational angular velocity.
		
		\subsection{Reinforcement learning algorithm with a maximum entropy objective}
		
		\paragraph{}
		The fish brain function is implemented using a deep reinforcement learning algorithm, specifically the soft actor-critic(SAC)algorithm was first proposed to address the poor training performance of DDPG on high-dimensional continuous control problems\cite{haarnojaSoftActorCriticAlgorithms2019}. Compared to the traditional optimization objective of reinforcement learning, SAC introduces the maximum entropy(ME) objective  to achieve a random policy. Due to its excellent performance in terms of training stability, sample efficiency, and convergence speed, SAC makes it possible to train directly using real samples collected by physical robots without the need for a simulation environment\cite{haarnojaSoftActorCriticAlgorithms2019}.
		
		\paragraph{}
		For traditional deep reinforcement learning algorithms, the learning objective is to learn policies that maximize the expected cumulative reward:
		
		\begin{equation}\label{equ_7}
			\pi^*=\arg\max_{\pi}\mathbb{E}_{(s_t,a_t)\sim\rho_\pi}\bigg[\sum\limits_t R\big(s_t,a_t\big)\bigg]
		\end{equation}
	
		\paragraph{}
		For reinforcement learning algorithms with a maximum entropy objective, the maximum entropy value for each output action is also required in addition to maximizing the expected cumulative reward, this is represented in equation \ref{equ_8}:
		
		\begin{equation}\label{equ_8}
			\pi^*=\arg\max_{\pi}\mathbb{E}_{(s_r,a_t)\sim\rho_\pi}[\sum_t\underbrace{R\big(s_t,a_t\big)}_{reward}+\alpha\underbrace{H\big(\pi\big(\mid s_t\big)\big)}_{entropy}]
		\end{equation}
	
		\paragraph{}
		where $\alpha$ is the temperature parameter that determines the relative importance of the entropy term versus the reward, which controls the stochasticity of the optimal policy.
		
		\paragraph{}
		The principle of maximum entropy is employed to formulate the stochastic policy, with the fundamental concept being to not exclude any potentially advantageous actions. In contrast to deep reinforcement learning algorithms utilizing deterministic policies, which tend to prioritize certain actions and discard others, those utilizing stochastic policies strive to distribute the probability of selecting any given action as evenly as possible, rather than focusing solely on a single action.
		
		During the policy evaluation step, the Q-value is iteratively updated in accordance with equation \ref{equ_9} until a state of convergence is reached:
		
		\begin{eqnarray}\label{equ_9}
			Q_{soft}^{\pi}\big(s_t,a_t\big)=r\big(s_t,a_t\big)+\gamma\mathbb{E}_{s_{t+1},a_{t+1}}\biggl[Q_{soft}^{\pi}\big(s_{t+1},a_{t+1}\big)-\alpha\log\bigl(\pi\big(a_{t+1}|\ s_{t+1}\bigr)\bigr)\biggr]
		\end{eqnarray}
	
		The policy improvement step involves updating the policy in accordance with equation \ref{equ_10}.
		
		\begin{eqnarray}\label{equ_10}
			\pi^{'}=\arg\min_{\pi_{k}\in I I}D_{K L}\left(\pi_{k}\left(\cdot| s_{t}\right)\|\dfrac{\exp\left(\dfrac{1}{\alpha}Q_{soft}^{\pi}\left(s_{t},\cdot\right)\right)}{Z_{soft}^{\pi}\left(s_{t}\right)}\right)
		\end{eqnarray}
	
		The two processes of policy evaluation and policy update are repeated in an iterative manner until the final policy $\pi$ converges to the optimal policy $\pi^{*}$ , which is defined as satisfying certain conditions:For $\forall \pi  \in \Pi $ and $\forall \left( {{{\rm{s}}_{\rm{t}}},{{\rm{a}}_{\rm{t}}}} \right) \in {\rm{S}} \times {\rm{A}}$, ${{\rm{Q}}^*}\left( {\;{{\rm{s}}_{\rm{t}}},{{\rm{a}}_{\rm{t}}}} \right) \ge {{\rm{Q}}^\pi }\left( {{{\rm{s}}_{\rm{t}}},{{\rm{a}}_{\rm{t}}}} \right)$.
		
		The pseudocode for the SAC algorithm is provided as follows.
		
		\begin{algorithm}
			\caption{Soft Actor-Critic\cite{haarnojaSoftActorCriticAlgorithms2019}}
			\label{alg:sac}
			\begin{algorithmic}
				\STATE{Hyper parameters:Q net1 $\theta_{1}$ ,Q net2 $\theta_{2}$ ,Policy-net $\phi$ }
				\STATE{Initialize target network weights:${\bar \theta _1} \leftarrow {\theta _1}$, ${\bar \theta _2} \leftarrow {\theta _2}$}
				\STATE{Initialize an empty replay pool:${\cal D} \leftarrow \emptyset $ }
				\FOR{each iteration step = 1:N}
					\FOR{each environment step = 1:T}
						\STATE{$\mathbf{a}_t\sim\pi_\phi\big(\mathbf{a}_t|\mathbf{s}_t\big)$}
						\STATE{$\mathbf{s}_{t+1}\sim p\big(\mathbf{s}_{t+1}\big|\mathbf{s}_t,\mathbf{a}_t\big)$}
						\STATE{$\mathcal{D}\leftarrow\mathcal{D}\cup\big\{\big(\mathbf{s}_t,\mathbf{a}_t,r\big(\mathbf{s}_t,\mathbf{a}_t\big),\mathbf{s}_{t+1}\big)\big\}$}
					\ENDFOR
					\FOR{gradient step = 1:n}
						\STATE{${\theta _i} \leftarrow {\theta _i} - {\lambda _Q}{\hat \nabla _{{\theta _i}}}{J_Q}\left( {{\theta _i}} \right){\rm{ for }}i \in \{ 1,2\} {\rm{ }}$}
						\STATE{$\phi  \leftarrow \phi  - {\lambda _\pi }{\hat \nabla _\phi }{J_\pi }(\phi )$}
						\STATE{$\alpha  \leftarrow \alpha  - \lambda {\hat \nabla _\alpha }J(\alpha )$}
						\STATE{${\bar \theta _i} \leftarrow \tau {\theta _i} + (1 - \tau ){\bar \theta _i}{\rm{ for }}i \in \{ 1,2\} $}
					\ENDFOR
				\ENDFOR
				\STATE{Optimized parameters:$\theta_{1}$,$\theta_{2}$,$\phi$}
			\end{algorithmic}
		\end{algorithm}
		
		\subsection{Modeling of agent fish kinematics parameters and power calculation}

		\paragraph{}
		In the natural world, the majority of fish possess bodies that resemble symmetrical airfoils. In this study,a NACA0012 airfoil is utilized for modeling the body of fish. To accurately simulate the movement of real fish, it is essential to regulate the undulations of the fish's body through the use of fish kinematics equations and to translate these fluctuations from the global Cartesian coordinate system to the local coordinate system of the fish's body motion. Excluding the direct influence of fins, the simplified wave curve of a Carangidae fish body can be represented as\cite{huangAUVAssistedDataGathering2020}:
		
		\begin{equation}\label{equ_11}
			Z({X_{fish}},t) = \left( {{c_1}\frac{{{X_{fish}}}}{L} + {c_2}{{\left( {\frac{{{X_{fish}}}}{L}} \right)}^2}} \right)A_{max}\sin \left( {k\frac{{{X_{fish}}}}{L} + wt} \right)L
		\end{equation}
	
		In equation \ref{equ_11}: $c_{1}$  and $c_{2}$  are the primary and secondary envelope coefficients;$X_{fish}$   is the distance from the current coordinate point to the head of the fish; \textit{L} is the length of the fish; \textit{k} is the wave number of the body stem curve, which can be calculated by $k = \frac{{2\pi }}{\lambda }$ ,the larger \textit{k} is, the more the fish tends to fluctuate forward, the smaller \textit{k} is, the more the fish tends to swing forward, $\lambda$ is the wavelength of the fish body, refer to the swimming pattern of the four major fish in the Yangtze River in China,  take \textit{k} is 3; $A_{max}$ is the max amplitude of fish body's wave.
	
		In the K$\acute{a}$rm$\acute{a}$n-swimming test, the energy consumption of fish swimming is taken into account. Given that the oscillation of the fish's body occurs perpendicular to the x-axis and along the y-axis of the local coordinate system, the dimensionless transient work ${w_t}^*$  and the total work of the fish's body  ${W_{\rm{t}}}^*$ in the current dimensionless time step can be calculated as:
		
		\begin{equation}\label{equ_12}
			{w_t}^{\rm{*}} = \oint\limits_s {{F_y}\delta y} ds
		\end{equation}
		
		\begin{equation}\label{equ_13}
			{W_{\rm{t}}}^*{\rm{ = }}\int\limits_0^t {{w_t}^*dt} 
		\end{equation}
	
		Where: ${w_t}^*$ is the transient work of the fish body at the current transient moment; ${W_t}^*$ is the total work from the starting moment to the current moment; \textit{s} is the closed curve formed by the two-dimensional surface of the fish body; $\delta y$  is the distance of the movement of the fish body mass along the y-axis of the accompanying coordinate system at the current moment; \textit{Fy} is the component of the combined force of the water flow on the fish body mass along the y-axis of the accompanying coordinate system at the current moment; for the local coordinate system of the fish body, the x-axis is the moving-forward direction of the fish body, and the y-axis is perpendicular to the moving-forward direction of the fish body.
		
		\subsection{Fluid solver validation}
		
		\paragraph{}
		Here, a validation case of fish swimming in the quiescent flow is used to verify the reliability of the fluid solver,the parameters are as follows: The translation and rotation are determined by Eqation. (5) and (6), the fish swims Reynolds number $Re = \rho {L^2}/T\mu  = 7000$, and the computational field is discreted by a 1000×500 scale cartesian grid with a total number of 500,000 grids, the initial position of the fish is [850,250],the validation case is implemented on a workstation equipped with an Nvidia RTX A4000 graphics card using Nvdia's CUDA technology for GPU parallel computing acceleration, the fish is depicted swimming in a straight line within the quiescent flow.
		
		\begin{figure}[!t]
			\centering
			\includegraphics[width=5in]{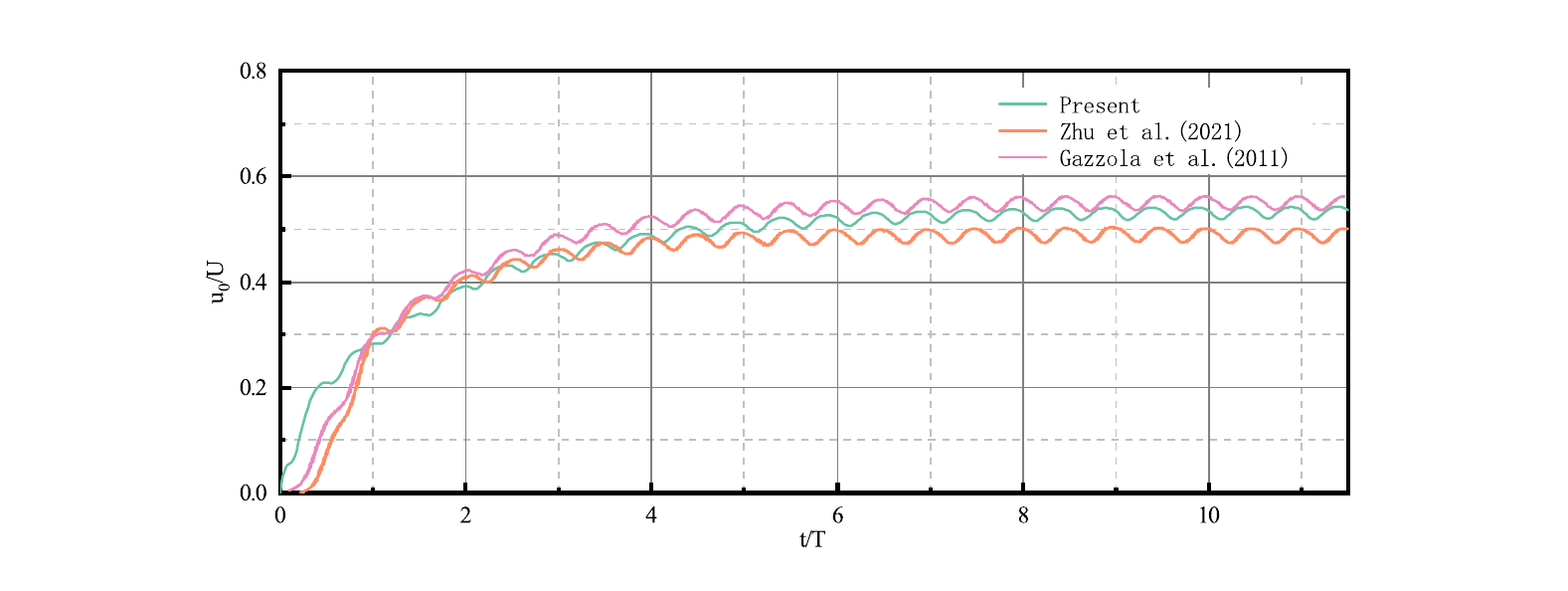}
			\caption{Time history graph of the forward swimming velocity of fish.}
			\label{fig_1_val}
		\end{figure}

		The time history of the forward swimming velocity of the fish is illustrated in figure \ref{fig_1_val}, where: $U{\rm{ = }}L/T$, the results are compared with Zhu and Gazzola. It is noteworthy that the maximum swimming speed in this study is slightly lower than that reported by Gazzola due to the variation in the initial body phase setting, however, this discrepancy does not affect the reinforcement learning decisions, and thus, it was not corrected in order to conserve computational resources.

		\section{Special model development}
		\subsection{Lateral-line machine development}
		\subsubsection{Lateral-line machine -knowledge acquisition}
		
\begin{figure}[!t]
			\centering
			\includegraphics[width=5in]{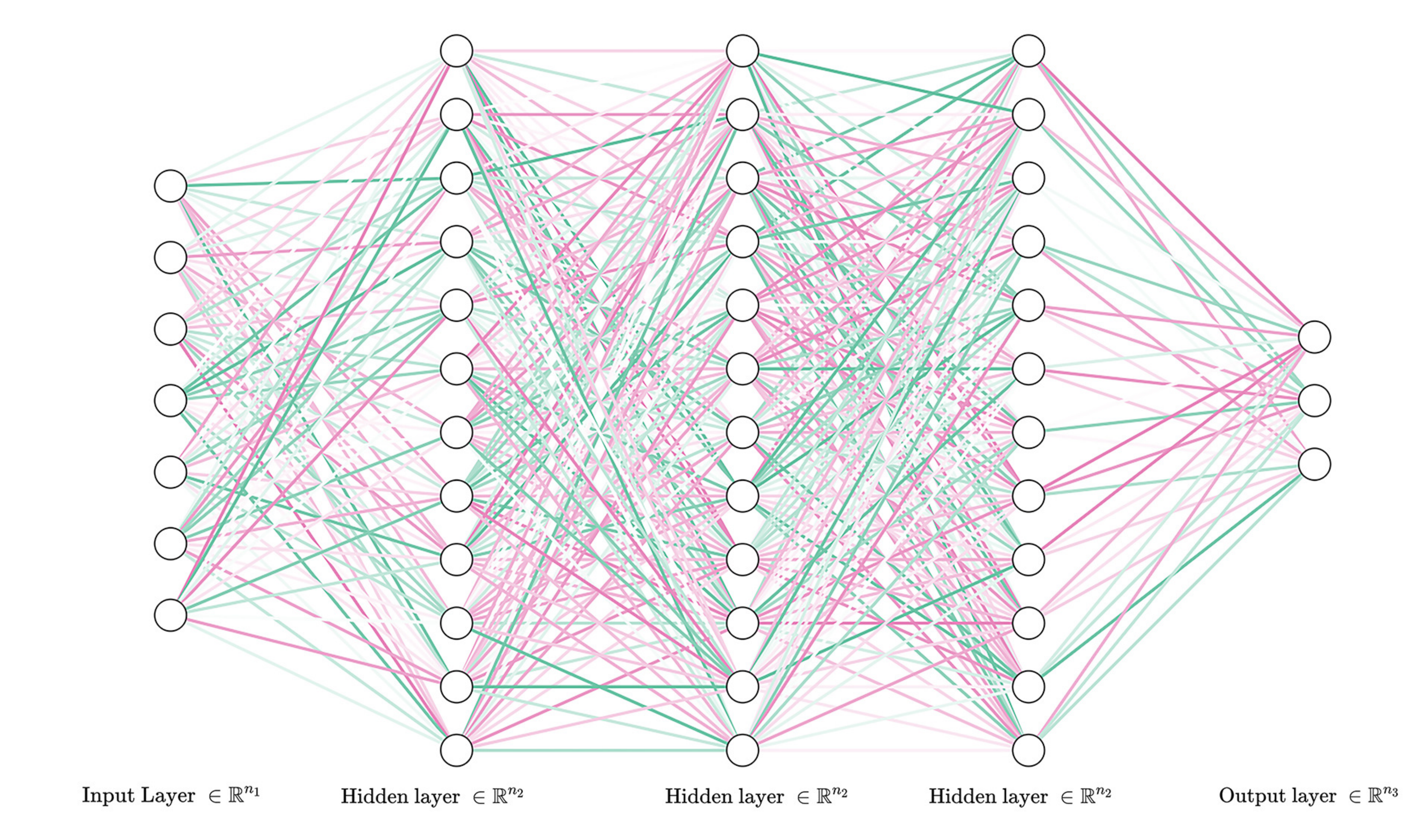}
			\caption{Diagram of the classification neural network.}
			\label{fig_2_dnn}
		\end{figure}

		\paragraph{}
		Fish have developed the ability to perceive the flow field over millions of years, and the lateral line enables fish to perceive their environment and perform complex maneuvers such as predatory swimming, reoaxis swimming, Kármán gait swimming, and fish-school swimming\cite{zhengArtificialLateralLine2017}. To enable the agent fish trained by deep reinforcement learning algorithms to perceive the flow field, it is not sufficient to implement end-to-end learning by simply inputting some feature parameters of the flow field into the state space. Due to the nonlinear nature of the flow field, the solution space in this natural problem is vast and cannot be exhausted. It would be unscientific to pass the raw flow field information to the neural network without any preprocessing and expect the neural network to automatically extract relevant information for perception and decision-making. Due to the inherent limitations of deep reinforcement learning algorithms in terms of learning efficiency, this approach would likely lead to difficulty in convergence of the algorithm and increase training time by orders of magnitude. Therefore, an additional lateral-line machine module is necessary to preprocess the flow field feature information used for policy decision, in order to reduce the solution space. 
		\paragraph{Method}
		The core of perceiving the flow field is to categorize the current perceived flow field based on the different flow field characteristic parameters perceived by the fish, for which a neural network for flow field classification needs to be constructed. This is a multi-classification problem in machine learning. The diagram of the classification neural network is shown in figure \ref{fig_2_dnn}.
		
		For multi-classification problems, the commonly used loss function is the cross-entropy loss\cite{hoRealWorldWeightCrossEntropyLoss2020}. The computation of the cross-entropy is presented in equation \ref{equ_14}.
		
		\begin{equation}\label{equ_14}
			{\mathop{\rm loss}\nolimits} (x,{\rm{ class }}) =  - \log \left( {\frac{{\exp (x[{\rm{ class }}]))}}{{\sum\limits_i {\exp } (x[i]}}} \right) =  - x[{\rm{ class }}] + \log \left( {\sum\limits_i {\exp } (x[i])} \right)
		\end{equation}
		
		Where: \textit{x} is the output of the last layer of the DNN neural network; \textit{class} is the index value corresponding to the upstream data type.
		
		Following the calculation of the loss function via cross-entropy, the adaptive moment optimizer (Adam) is employed to optimize the loss function and execute gradient backpropagation. This step is utilized to optimize the neural network for classification.
		
		\subsubsection{Training process of the lateral-line machine module}
		
\begin{figure}[!t]
			\centering
			\includegraphics[width=5in]{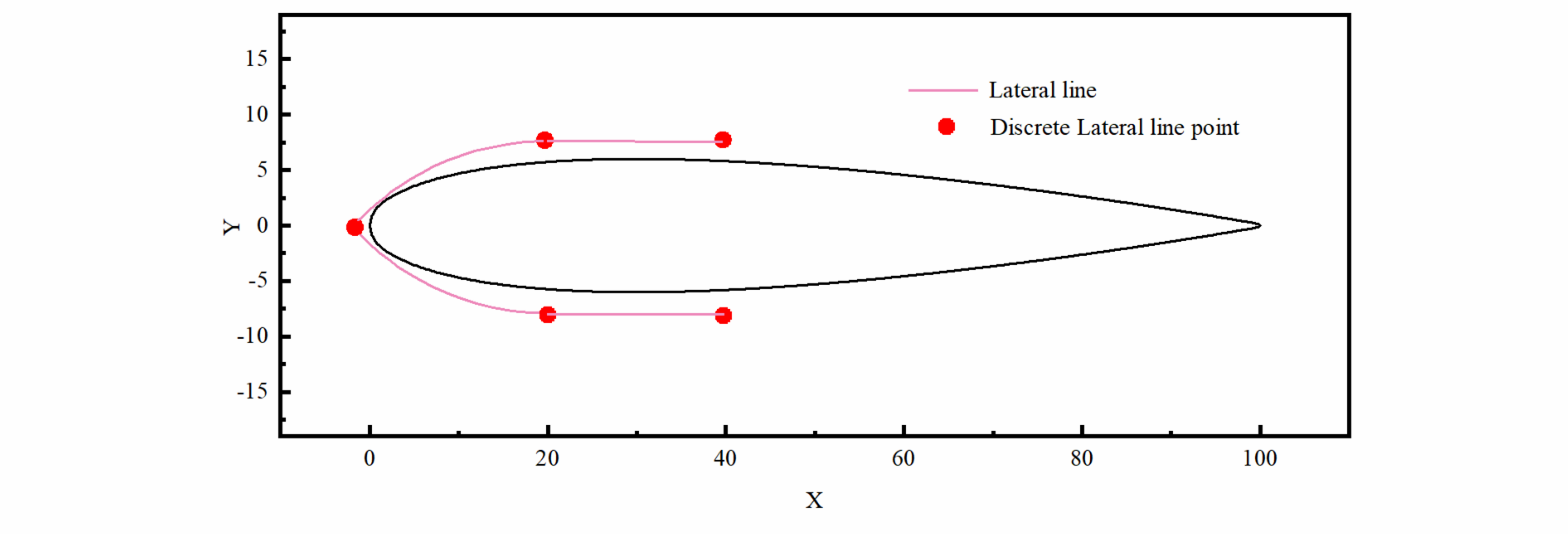}
			\caption{Diagram of agent fish's lateral-line monitoring points}
			\label{fig_3_LL}
		\end{figure}

		\paragraph{}
		The dataset for the flow field classification training was acquired from the lateral line perception data collected during simulations of fish swimming in various flow fields, with a total of 6000 data points. The layout of the lateral line is illustrated in figure \ref{fig_3_LL}. As the lateral line of a real fish body comprises a vast number of cellular neural thalami, it is infeasible to fully abstract the flow field information perceived by the neural thalami of the lateral line in mathematical terms, taking into account that the head of the fish is stimulated by the water flow when swimming\cite{liaoRoleLateralLine2006}. To optimize computational efficiency while effectively capturing the flow field information around the fish's body, the continuous lateral line was discretized into five monitoring points and then averaged as the observed signal of the whole fish's body for the local flow field.
		
		The training data is represented by multiple 1-dimensional vectors, the format of the data is as follows: $[t,{x_{grid}},{y_{grid}},{u_x},{u_y},P,T]$.Where \textit{t} represents the current flow field solution time; \textit{x} grid denotes the horizontal coordinate of the lateral grid point; \textit{y} grid denotes the vertical coordinate of the lateral grid point; $u_{x}$ represents the x-direction flow velocity at the lateral grid point; $u_{y}$ represents the y-direction flow velocity at the lateral grid point; \textit{P} represents the flow field pressure at the lateral grid point; \textit{T} represents the total flow velocity magnitude at the lateral grid point, $T = \sqrt {{u_x}^2 + {u_y}^2} $.

\begin{figure}[H]
			\centering  
			\subfigbottomskip=2pt 
			\subfigcapskip=-5pt 
			\subfigure[Training accuracy of lateral-line machine]{
				\includegraphics[width=0.48\linewidth]{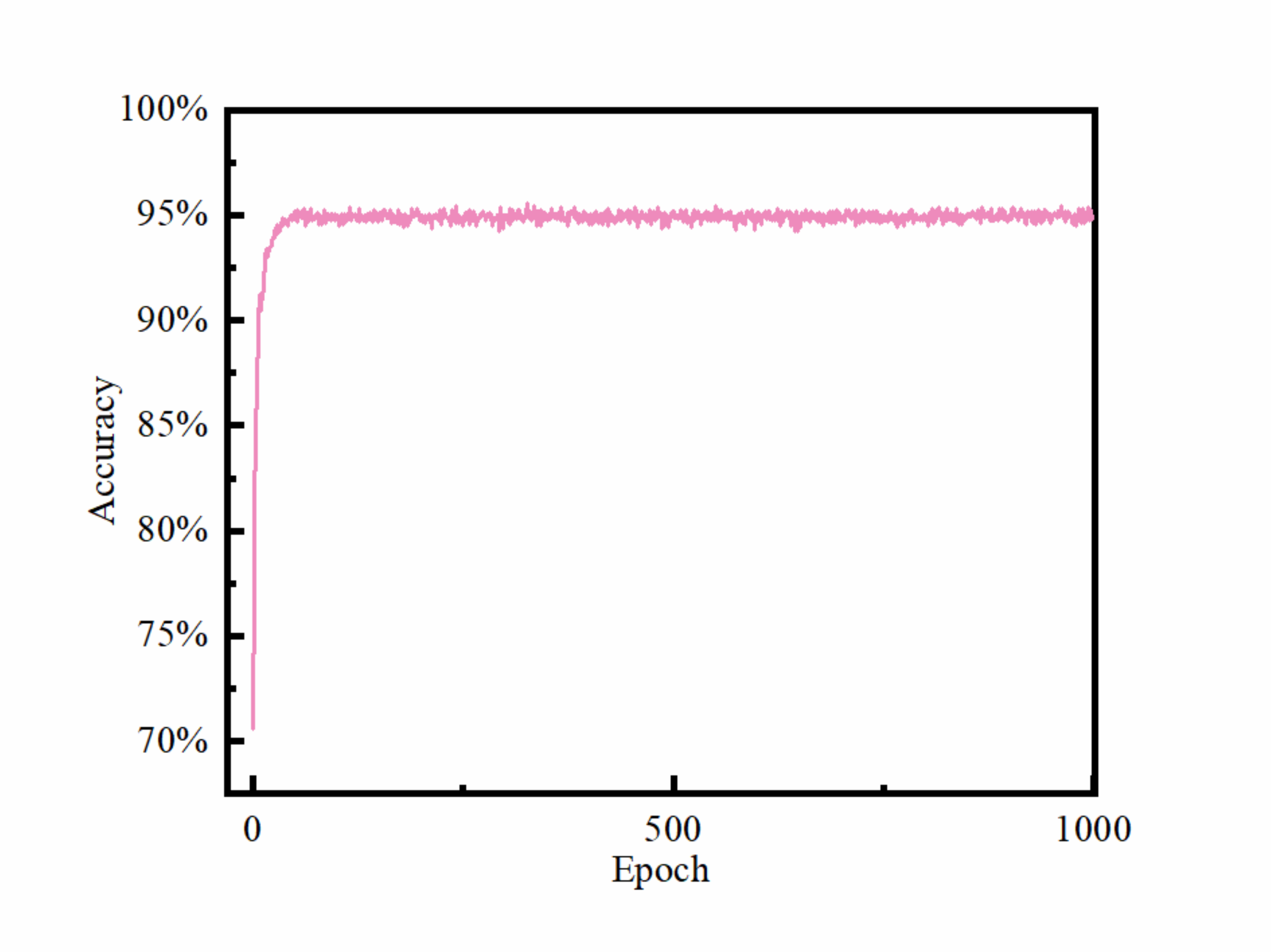}}
			\subfigure[Training loss of lateral-line machine]{
				\includegraphics[width=0.48\linewidth]{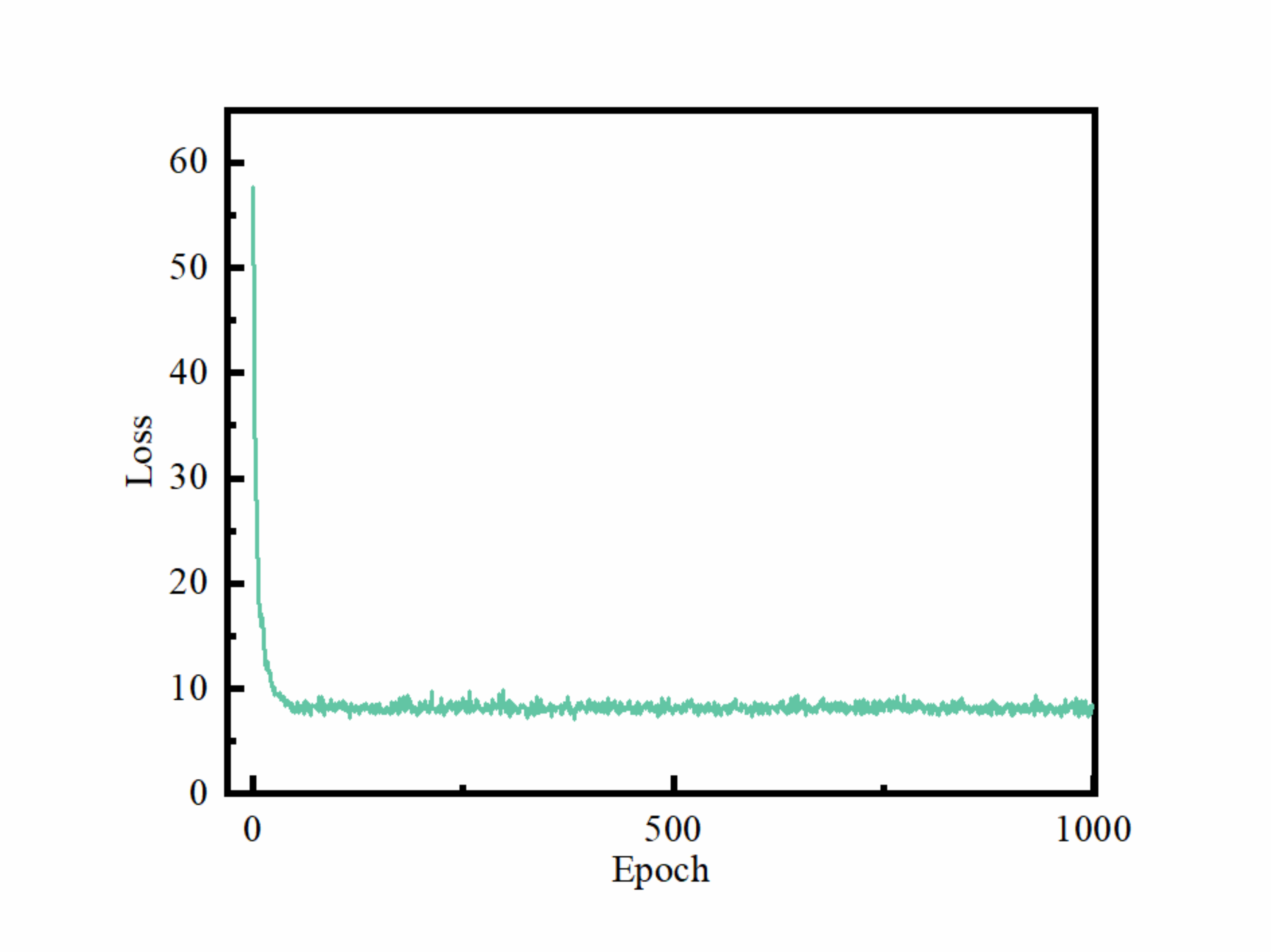}}
			\caption{Graph of training result }
			\label{fig_4_LL_result}
		\end{figure}

		After obtaining the collected lateral line perception dataset in the specified format, the lateral line perception machine can be trained using the labeled data. The learning rate for the lateral-line machine is set to 0.01, the batch size is set to 64 and overall, the training is done for 1000 epochs. After each round of training, the dataset  $\mathbb{X}$ is used as input and the classification of the flow field is predicted by the lateral-line machine to evaluate the training effect. The training accuracy is shown in figure \ref{fig_4_LL_result} (a)and the training loss is shown in figure \ref{fig_4_LL_result} (b). It can be observed that after a certain number of training epochs, the accuracy of the dataset increases from 70\% to 95\%, and this accuracy meets the prediction requirements for the classification of the flow field, thus, the trained lateral-line machine can be utilized for the classification of the flow field.

		\subsection{The development of macro-action system}
		
		\paragraph{}
		When designing the reinforcement learning action space, previous researchers often employed a single form of elemental action\cite{zhuNumericalStudyFish2021,yanNumericalSimulationMethod2020}. If the trained agent fish enters an unfamiliar environment, the learned swimming policy will be partially invalid and will not be well transferred, resulting in poor generalization of the trained knowledge. To address this, it is necessary to design a coupled knowledge transfer system to enable the transfer of trained knowledge. Since there is a function mapping relationship between the vortex shedding frequency and the fish's optimal waving frequency in vortex flow fields with different characteristic frequencies. Firstly, the type of the current flow field can be roughly determined by the lateral-line machine, then the optimal waving frequency matrix suitable for the current flow field can be selected from the existing waving frequency repository, and macro-action can be performed in this waving frequency matrix. The previously trained swimming policy can adapt to a new flow field, achieving knowledge transfer process. The following section describes the implementation of the macro-action system.
		
		The design of action space is often based on a series of basic elemental actions. For example, in the design of underwater AUV robots, the robot's swimming often involves the linkage of multiple tail joints. If the basic action space is designed using basic elemental motions, then more advanced actions are difficult to achieve, such as C-maneuvers, burst and coast, among others. Another example is in video games, such as FIFA, where humans can manipulate the virtual football player to execute advanced techniques like Marseille turn and Panenka penalty based on their own experience, but it is challenging for an agent to learn and perform such advanced movements in the game based on elemental action expansion. This is due to the fact that it is extremely difficult for the agent to encounter and learn these advanced actions on its own without a corresponding, targeted reward function design.
		
		For live fish, the waving frequency cannot be constant, and the combination of many complex waving frequencies forms three macro-actions: cruise, acceleration, and deceleration. For example, if a fish moves at the waving frequency  $\omega_{1}$, there are three kinds of maneuvers it can choose: 1.Continue to move at the waving frequency  $\omega_{1}$, the speed of the fish will not change, which is the cruising maneuver; 2.The fish increases the waving frequency from   $\omega_{1}$ to  $\omega_{0}$, the swimming speed of the fish will also increases, which is the acceleration maneuver; 3.The fish increases the waving frequency from  $\omega_{1}$ to  $\omega_{2}$ , the swimming speed of the fish will also decreases, which is the deceleration maneuver.
		
		Therefore, drawing inspiration from the motion of real fish in nature, the macro-action system is set as $[{a_{{\rm{accelerate}}}},{a_{{\rm{cruise}}}},{a_{{\rm{decelerate}}}}]$, where: ${a_{{\rm{accelerate}}}}$represents the acceleration maneuver, meaning the agent fish will increase the current waving frequency to achieve acceleration; ${a_{{\rm{cruise}}}}$ represents the cruising maneuver, meaning the agent fish will maintain the current waving frequency and continue to cruise at a uniform speed;${a_{{\rm{decelerate}}}}$ represents the deceleration maneuver, meaning the agent fish will decrease the waving frequency to achieve deceleration. In order to prevent the agent fish from making irrational actions, it is necessary to ban actions that are not allowed by the rules of nature in a specific state and would result in severe consequences. For example, when the agent fish has accelerated to the maximum waving frequency set by the system, selecting acceleration action again would be invalid because real fish in nature cannot indefinitely accelerate and must have a maximum waving frequency, the same applies to deceleration maneuvers. Here, we design an illegal action banning system by combining specific parameters of the state space to prohibit illegal actions in a specific state.
		
		\begin{enumerate}
			\item When the agent fish recognizes ${\omega _t}{\rm{ = }}{\omega _{\max }}$, the acceleration maneuver is banned and the action space becomes $[0,{a_{{\rm{cruise}}}},{a_{{\rm{decelerate}}}}]$, and the agent fish will only choose cruise and deceleration.
			\item When the agent fish recognizes ${\omega _t}{\rm{ = }}{\omega _{\min }}$, the deceleration maneuver is banned and the action space is changed to $[{a_{{\rm{accelerate}}}},{a_{{\rm{cruise}}}},0]$, and the agent fish will only choose cruise and acceleration.
			\item If $\omega_{t}$ does not belong to previous condition, the action space is $[{a_{{\rm{accelerate}}}},{a_{{\rm{cruise}}}},{a_{{\rm{decelerate}}}}]$ , and the agent fish can take three actions: acceleration, cruise and deceleration.
		\end{enumerate}

		\subsection{Coupled Logic Framework diagram}
		
\begin{figure}[!t]
			\centering
			\includegraphics[width=5in]{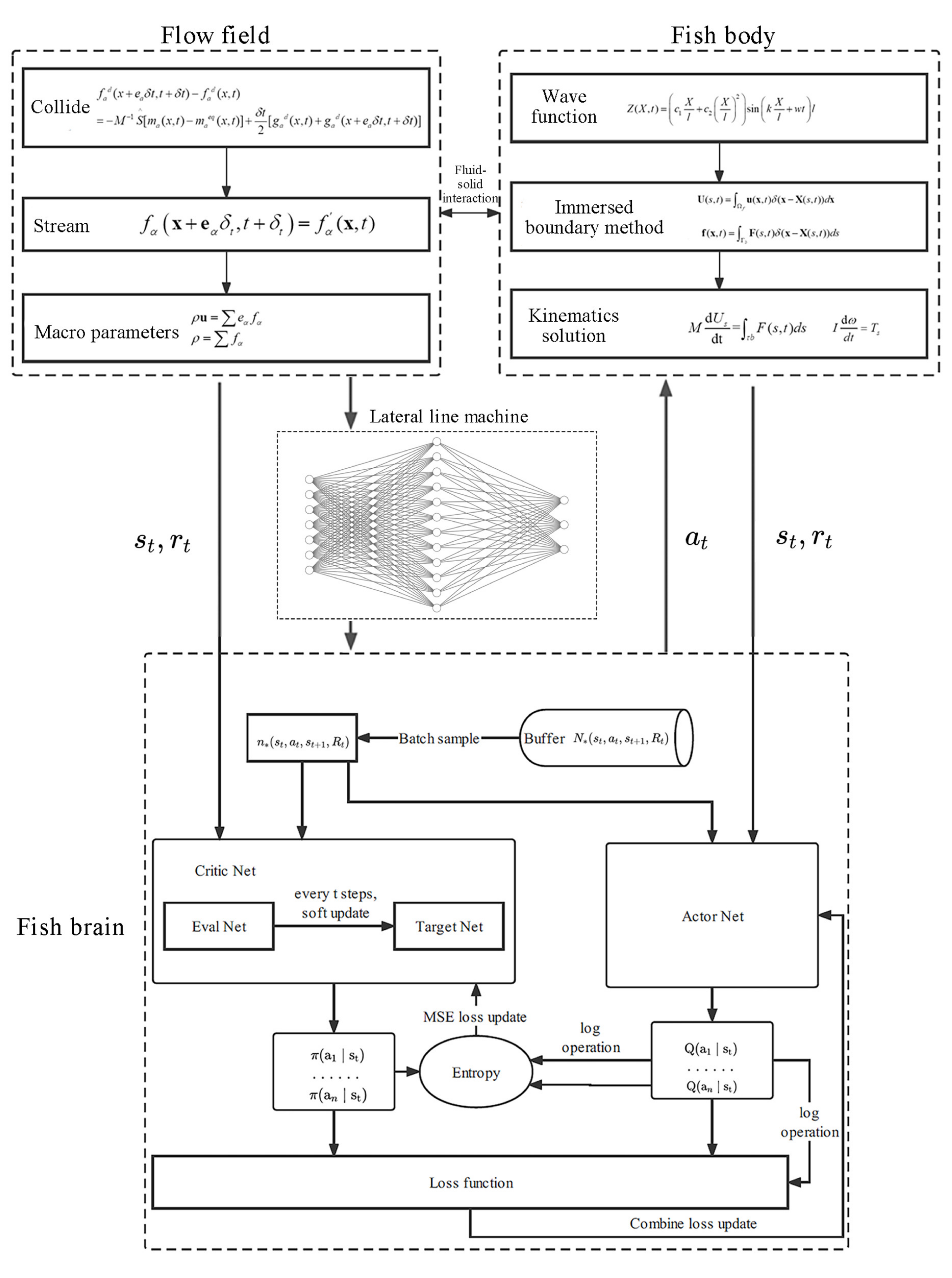}
			\caption{Diagram of the coupled computing platform}
			\label{fig_5_coupled}
		\end{figure}

		\paragraph{}
		The diagram of the coupled computing platform is shown in figure  \ref{fig_5_coupled}.The entire coupled system consists of a fish brain module, a lateral-line machine module, a fish body module and a flow field module, where the flow field habitat module provides the living space for the fish, the fish brain module is responsible for performing environmental perception and decision making functions, and the fish body module is responsible for receiving commands from the brain module. An object-oriented integration framework was developed by integrating all these modules into the final simulator. The role of each module is described as follows: 
		
		\begin{enumerate}
			\item The agent fish, which combines a lateral-line machine and deep reinforcement learning, is placed in the flow field to presense the flow field;
			\item The Fish brain module receives the state $s_{t}$ from the habitat side of the flow field, inputs the current state $s_{t}$ into the current policy value estimation model based on the SAC algorithm, and outputs the probability density function of the set of all actions, wherein the current policy value estimation model refers to the SAC deep reinforcement learning algorithm based on the;
			\item Randomly selecting an action based on the probability distribution ${\bf{P}}\{ a = {a_i}\} (i = 1,2,...)$ of all actions of the set of actions and using the action as the current target action;
			\item Generating control instructions for executing said current target action and transmitting said control instructions to fish body to execute $a_{t}$ in the flow field;
			\item Receiving new state $s_{t+1}$ and reward $R_{t}$ from flow field habitat ,which are calulated by interaction between flow field habitat and fish swimming;
			\item The current state $s_{t}$,the new state $s_{t+1}$,the action $a_{t}$ and reward $R_{t}$ are packed as an experience sample into a replay pool $D$  in a tuple format as follows:$\left[ {{s_t},{a_t},{s_{t + 1}},{R_t}} \right]$;
			\item Retrieving a sereis of batch experience samples ${n_*}\left( {{s_t},{a_t},{s_{t + 1}},{R_t}} \right)$ from replay pool and training strategic value estimation model based on experience samples ${n_*}\left( {{s_t},{a_t},{s_{t + 1}},{R_t}} \right)$ to obtain a new strategic value estimation model;
			\item Using new state information as current state information, and Return Step2 with new policy value estimation model. 
		\end{enumerate}
	
		\section{Results}
		
		\subsection{Task1-Point to point predation swimming}
		
		\paragraph{}
		The performance of our platform is initially evaluated using a point-to-point predation swimming example. In this training condition, the food's position is altered while the fish is swimming, providing a good verification of the agent fish's ability to capture a moving target.
		
		\subsubsection{Computational parameters}
		
		1.Hyper parameters of Deep reinforcement learning module
		
		When fish swim in the water, their visual system provides a navigation function for fish movement in the water. In order to make the intelligent body have certain visual function and make the intelligent body simulate the predatory movement of fish as realistically as possible, the observation state is set as a one-dimensional vector $[{X_{dis}},{Y_{dis}},{S_{dis}},{\theta _{food}}]$, where: ${X_{{\rm{dis}}}}$ is the x-direction distance of the fish head from the food, ${Y_{{\rm{dis}}}}$ is the y-direction distance of the fish head from the food, ${S_{dis}}$ is the absolute distance, ${\theta _{food}}$ is the directional angle of the fish body relative to the food, and the lateral-line machine is banned because the swimming task is simple and does not require the perception of complex flow field information. At the same time, in order to give the agent fish the ability to capture time sequence, the state information of several recent steps are superimposed in sequential order as the input information of the current moment, and the assembled state tuple is shown in equation \ref{equ_15}, and the stacked vectors are input into the deep neural network, which can make the intelligent body have short-term memory capability, and according to the experiment of Zhu\cite{zhuNumericalStudyFish2021}.It only needs to stack the last four waving periods’s state to satisfy the agent's requirement for time sequence information. The action space consists of three orders: $\left[ { - 1,0, + 1} \right]$, [-1] means that the fish rotates clockwise by 15° on the axis of the center of mass, [0] means that no effect saction is taken, and [+1] means that the fish rotates counterclockwise by 15° on the axis of the center of mass.
		
		\begin{equation}\label{equ_15}
			\left[ {\begin{array}{*{20}{c}}
					{{{[{X_{dis}},{Y_{dis}},{S_{dis}},{\theta _{food}}]}_{\rm{n}}}}\\
					{{{[{X_{dis}},{Y_{dis}},{S_{dis}},{\theta _{food}}]}_{n - 1}}}\\
					{{{[{X_{dis}},{Y_{dis}},{S_{dis}},{\theta _{food}}]}_{n - 2}}}\\
					{{{[{X_{dis}},{Y_{dis}},{S_{dis}},{\theta _{food}}]}_{n - 3}}}
			\end{array}} \right]
		\end{equation}
	
		When the fish brain receives the state matrix from the environment, the probability distribution matrix $[{P_a}_1,{P_a}_2,{P_a}_3.......{P_{an}}]$  corresponding to the action space matrix $[{a_1},{a_2},{a_3}......{a_n}]$ in the current state can be analyzed after perception and decision-making by the deep reinforcement learning module, and then the action can be selected based on this matrix. The action that is more likely to result in the agent obtaining more rewards in the current environment will have a higher probability of being selected.
		
		The effectiveness of the reward setting is directly related to whether the agent fish can accomplish its goal as expected. The reward function also acts as a bridge between the agent fish and the environment. Given the specific and complex nature of the flow field environment, the reward for the reinforcement learning module is set as follows: if the navigation angle or distance between the agent fish and the food gradually decreases, indicating that the fish is on the correct path, the reward is +1, otherwise, the reward is -1. If the agent fish successfully obtains the food, a larger positive reward of +20 is given. If the agent fish hits the wall or successfully eats the food, the episode ends.
		
		The hyperparameters of the deep reinforcement learning module are as follows: Policy loss entropy temperature $\alpha$ is performed by adaptive computation\cite{haarnojaSoftActorCriticAlgorithms2019}, the learning rate is 0.0005, the time discount factor $\gamma$ is 0.99,the sample’s batch size is 64, and the network soft update parameters $\tau$ is 0.01. The total number of layers of the deep neural network is 4: the number of nodes in the input layer is 16; the number of nodes in the hidden layer is 2, the number of nodes in each layer is 256, and the number of nodes in the output layer is 3. Additionally, the Softmax classification layer is employed to compress the final output probability distribution function of the action space. To improve the generalization of the training results, the agent fish is trained with random initial positions of the fish body. This approach ensures that the trained agent fish has the ability to capture moving targets at any positions, with the x-coordinates of the fish body being any random integer between [500, 750] and the y-coordinates of the fish body being any random integer between [25, 475].\\
		2.Hyper parameters of fluid solver
		
		For the fluid solver module, all physical units are converted into dimensionless lattice units for the convenience of presenting the calculation results. Characters with the "*" superscript denote dimensionless parameters in the experiment. A 1000×500 scale Cartesian grid is used to discretize the computational field, with a total of 500,000 grids. The waving period \textit{T} of the fish body is chosen as the characteristic time of the flow field, and in this calculation example, the waving period \textit{T} of the fish body is fixed as 600 dimensionless lattice time. The speed of the food's movement is ${u^*}{\rm{ = }}0.3L/T$. This case is implemented on a workstation configured with an Nvidia RTX A4000 graphics card, utilizing Nvidia's CUDA technology for GPU parallel computing.
		
		\subsubsection{Numerical simulation results}
		
\begin{figure}[!t]
			\centering
			\includegraphics[width=5in]{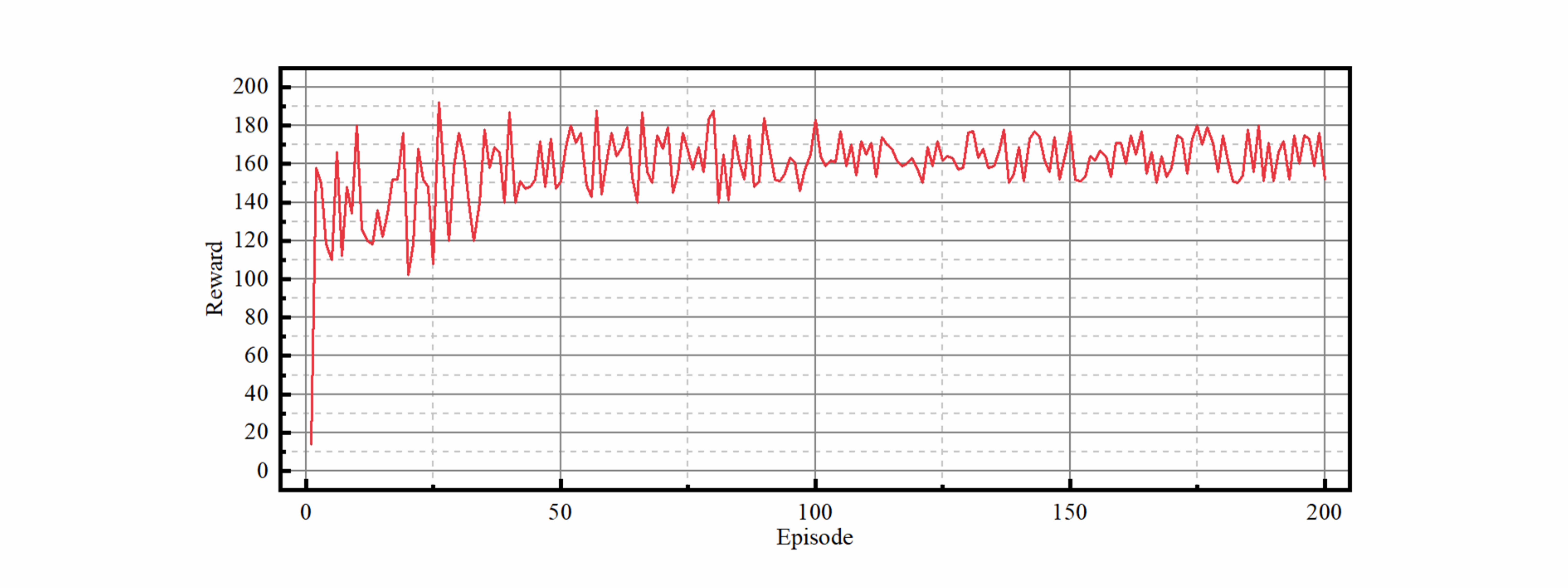}
			\caption{Reward received by the agent fish in predation swimming}
			\label{fig_6_reward1}
		\end{figure}

\begin{figure}[!t]
			\centering
			\includegraphics[width=5in]{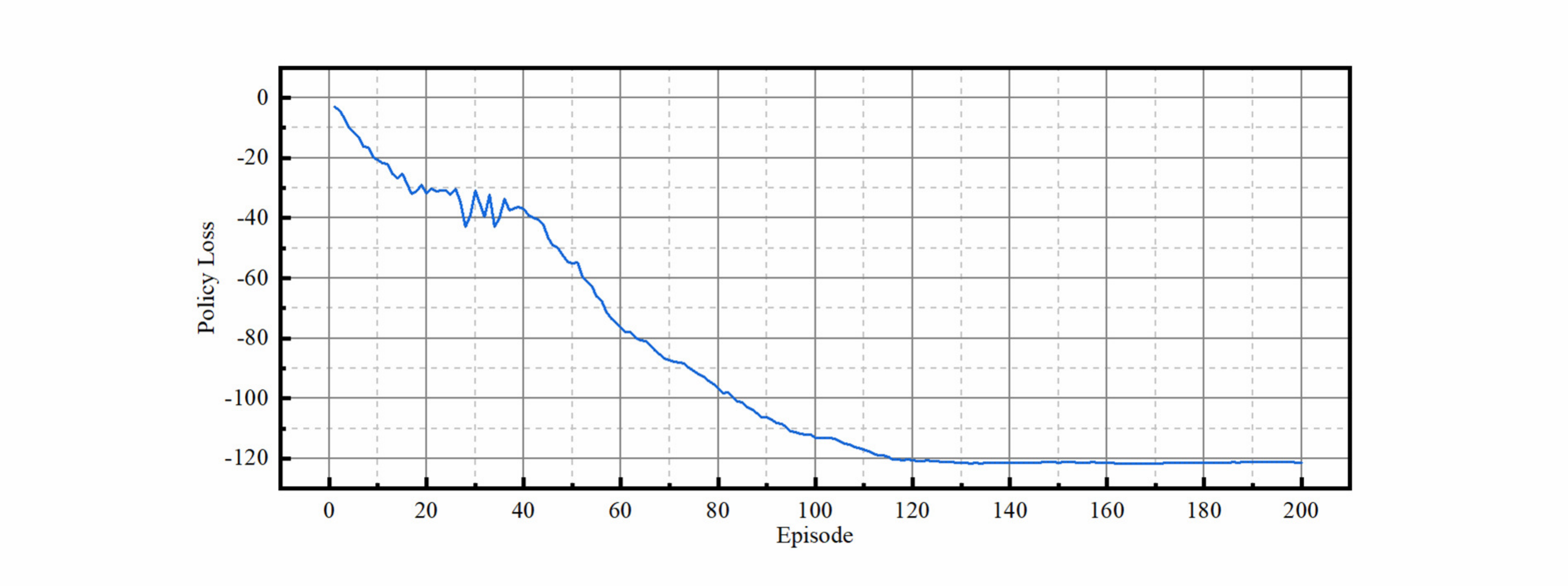}
			\caption{Loss of actor policy network in predation swimming}
			\label{fig_7_loss1}
		\end{figure}

		\paragraph{}
		Figure \ref{fig_6_reward1} illustrates the rewards received by the agent fish. It can be observed that the fish attained a relatively high level of reward after only a few episodes of training. This is a significant improvement when compared to the value-based reinforcement learning algorithm used in previous studies\cite{zhuNumericalStudyFish2021}. The SAC algorithm, which is based on the maximum entropy objective and random policy, utilized in this study has notable advantages in terms of convergence speed and training efficiency. After 35 episodes, the rewards obtained by the fish have stabilized, indicating that the policy network has effectively learned the correct knowledge of the capture motion food swimming policy through self-optimization. As a result, the fish is able to obtain higher rewards in the environment of food movement. It is worth noting that the rewards obtained may exhibit slight deviations due to the varying initial positions of the fish. Ultimately, whether the fish reaches the food is the criterion used to determine the success of the training. Figure \ref{fig_7_loss1} illustrates the loss of the actor policy network. The policy error exhibits a smooth decreasing curve, indicating that the fish has reached its optimal swimming policy in this test. The loss stabilizes around 120 episodes, signaling the completion of policy optimization and convergence.
		
\begin{figure}[!t]
			\centering
			\includegraphics[width=5in]{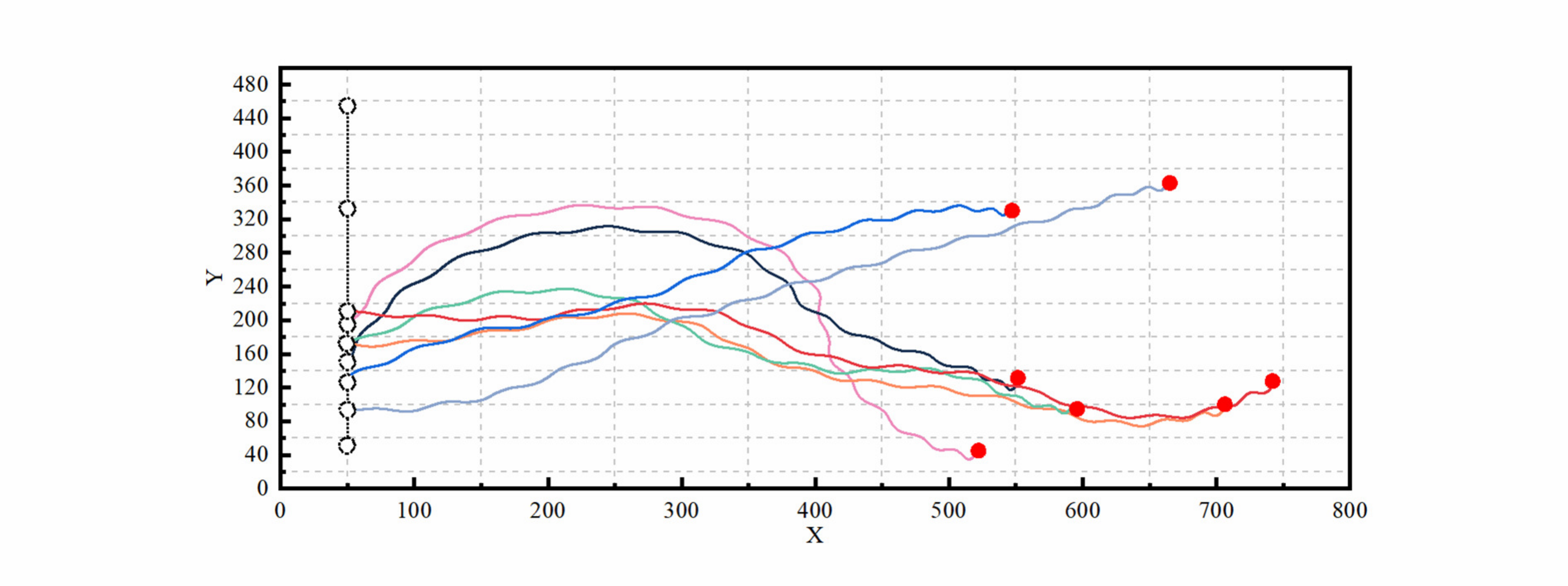}
			\caption{The trajectory graph of the fish at different initial positions}
			\label{fig_8_locus1}
		\end{figure}

		Figure \ref{fig_8_locus1} illustrates the trajectory of the fish at different initial positions when the target is in motion. It is noteworthy that the food in the left flow field is constantly in back-and-forth movement. As the fish becomes aware of the target's movement, it continually assesses its position relative to the food in real-time and adjusts its motion accordingly, in order to obtain the maximum reward in the current environment, while simultaneously ensuring it is on the shortest path to the food. This ultimately leads to the successful attainment of the target point and the food. The presence of a moving target results in a more complex and expansive solution space in comparison to a stationary target. However, the fish is still able to successfully complete the task, highlighting the reliability and robustness of the computing platform.
		
\begin{figure}[H]
			\centering  
			\subfigbottomskip=2pt
			\subfigcapskip=-5pt 
			\subfigure[$t^{*}=5.0T$]{
				\includegraphics[width=0.48\linewidth]{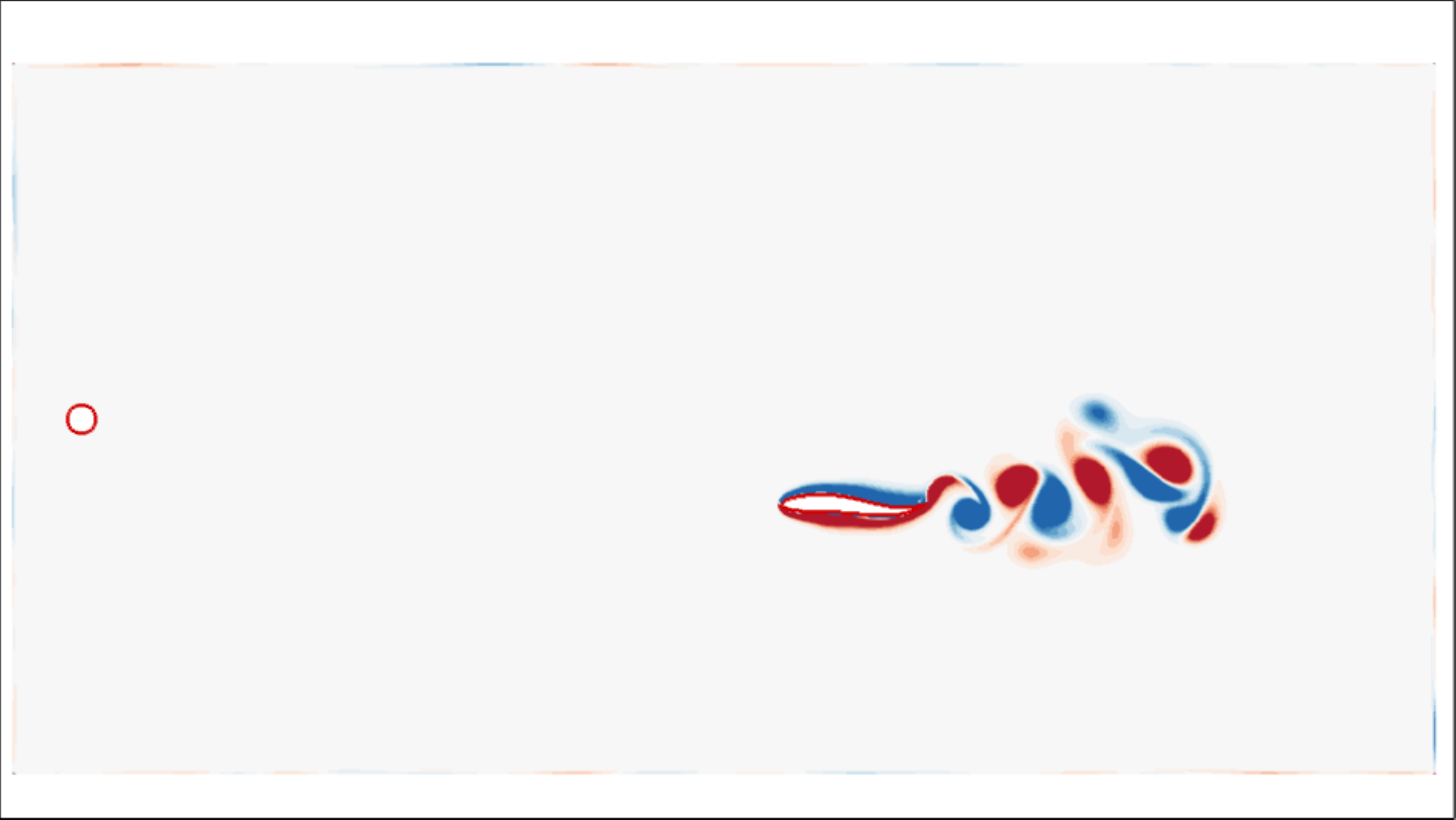}}
			\subfigure[$t^{*}=8.3T$]{
				\includegraphics[width=0.48\linewidth]{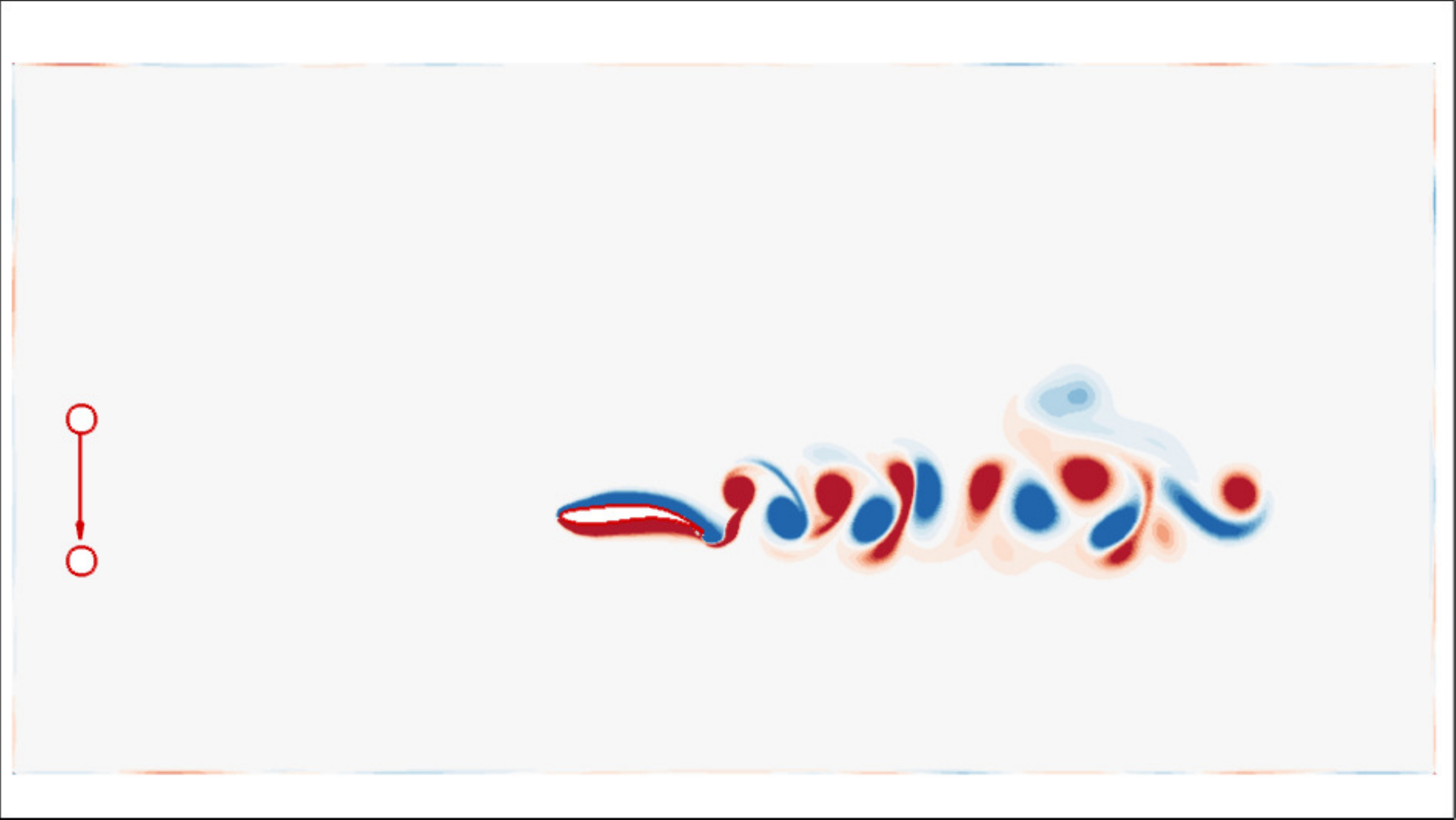}}
			\\
			\subfigure[$t^{*}=10.0T$]{
				\includegraphics[width=0.48\linewidth]{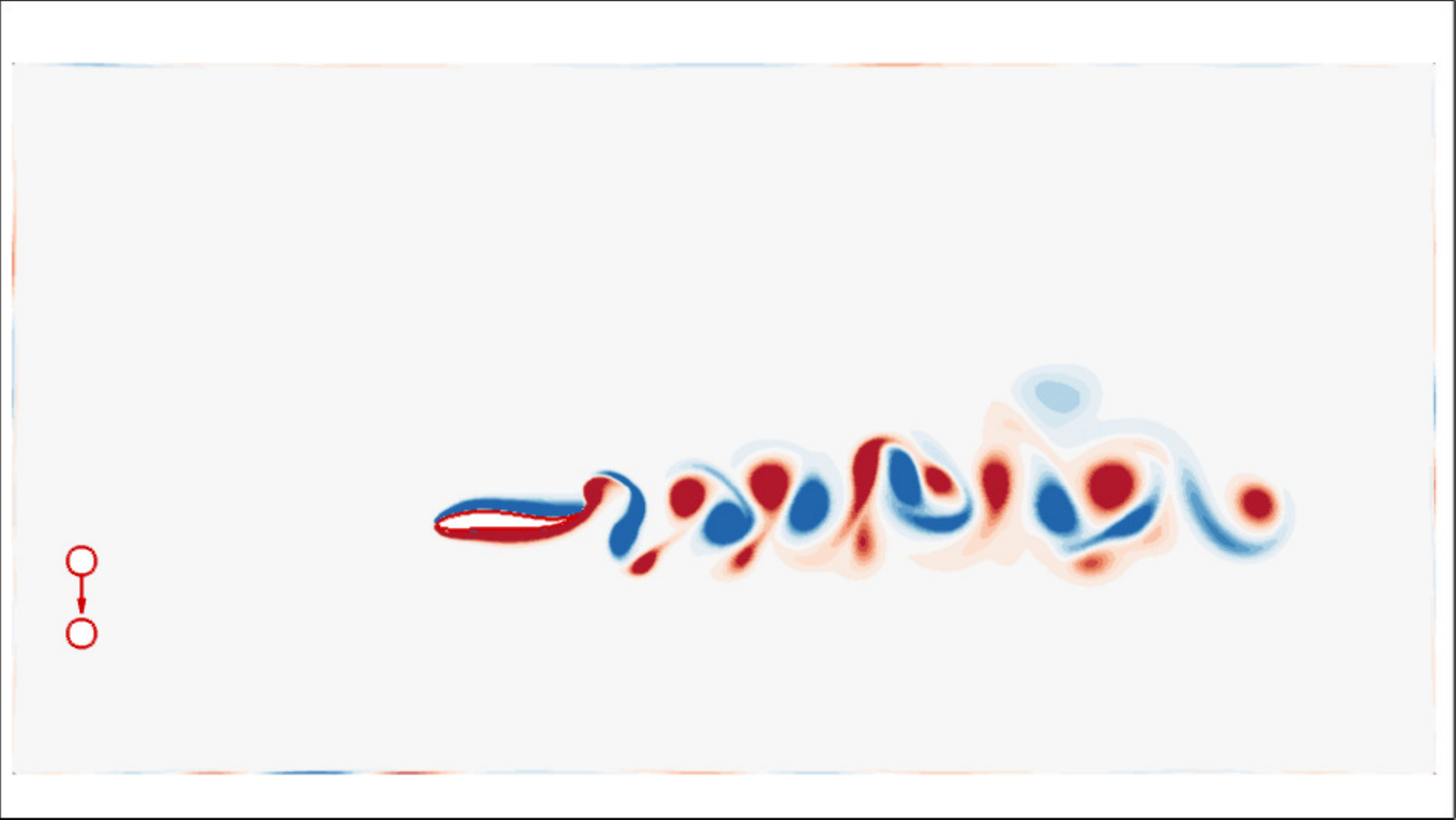}}
			\subfigure[$t^{*}=11.6T$]{
				\includegraphics[width=0.48\linewidth]{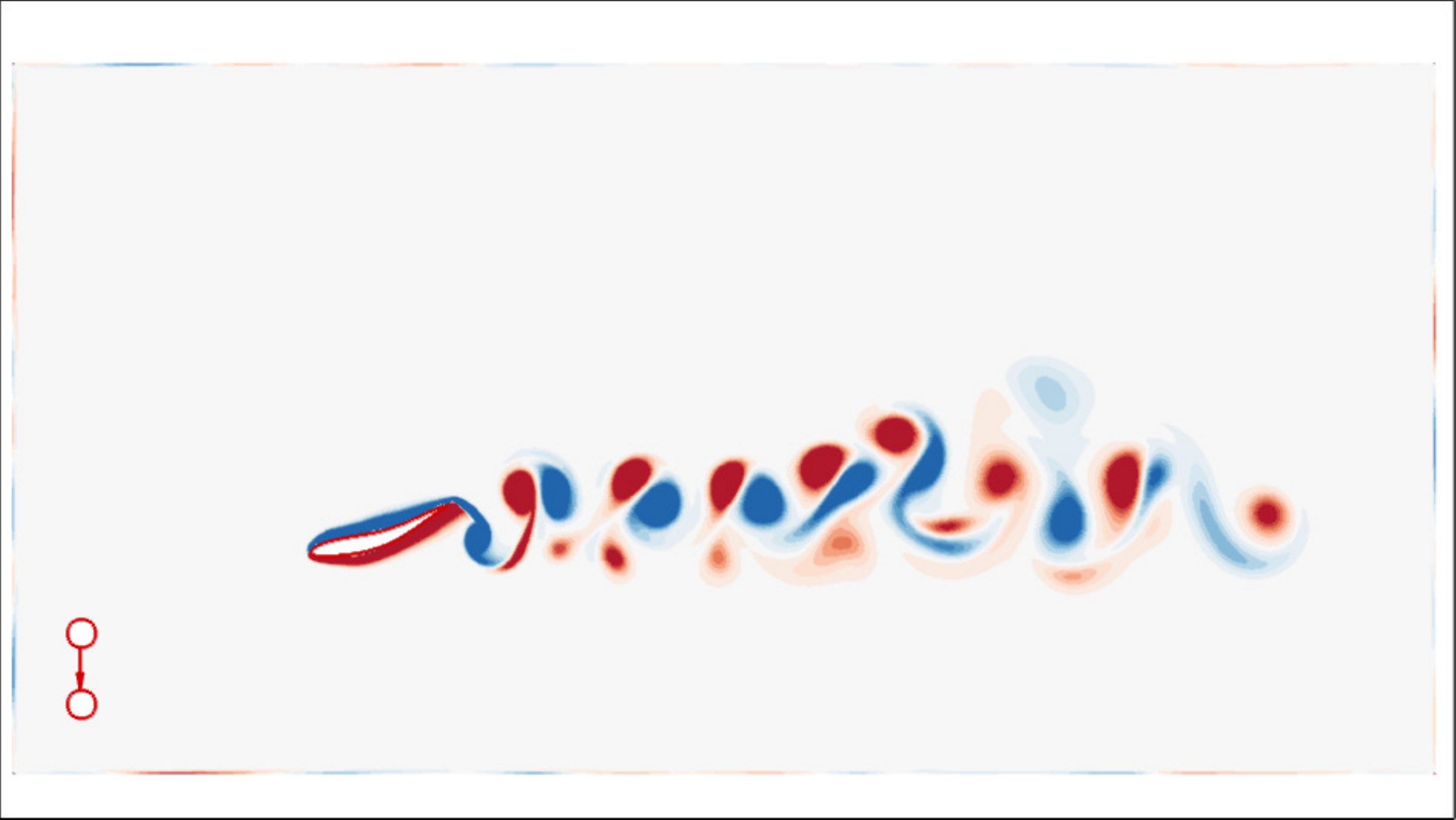}}
			\\
			\subfigure[$t^{*}=13.3T$]{
				\includegraphics[width=0.48\linewidth]{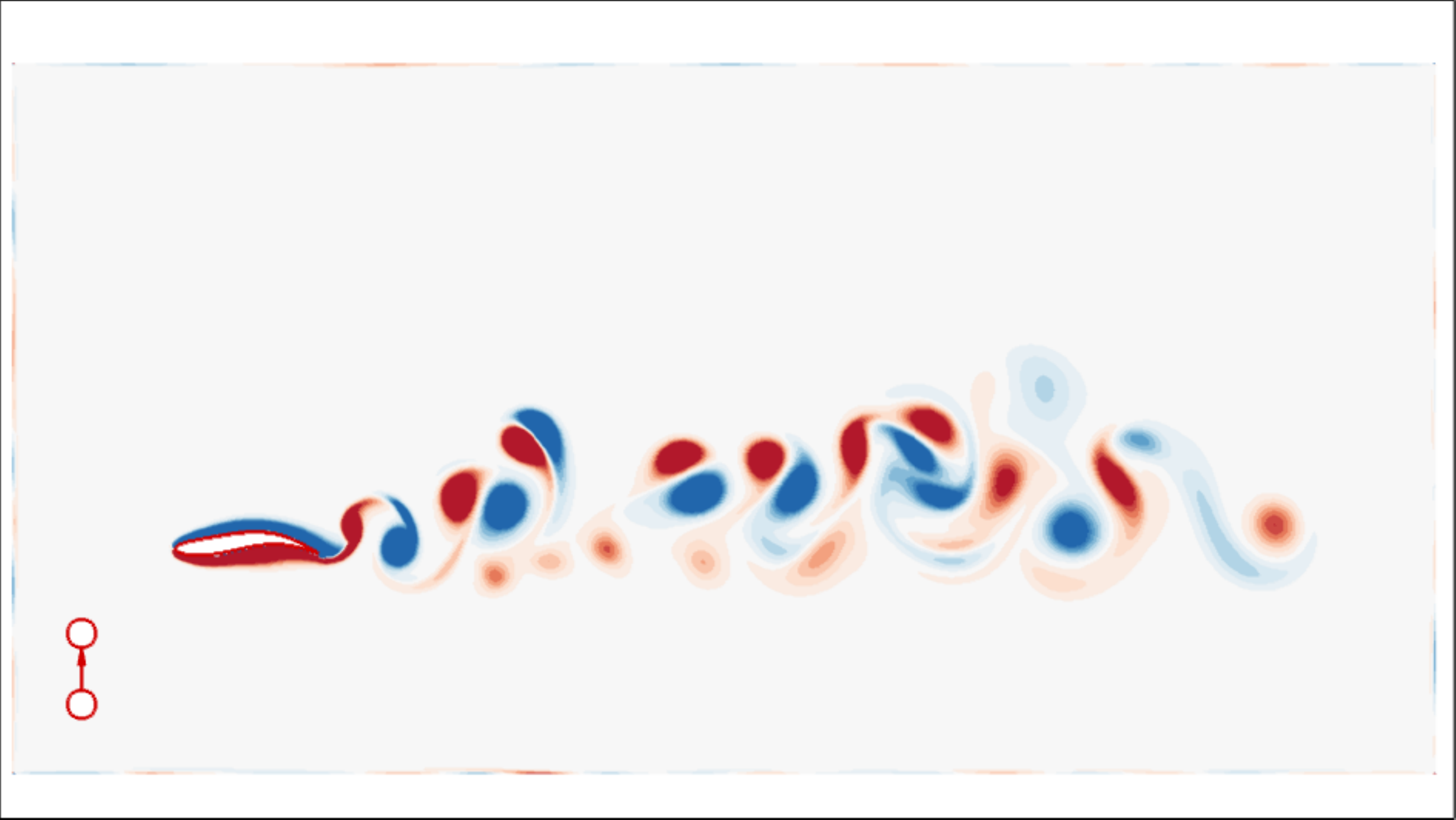}}
			\subfigure[$t^{*}=14.3T$]{
				\includegraphics[width=0.48\linewidth]{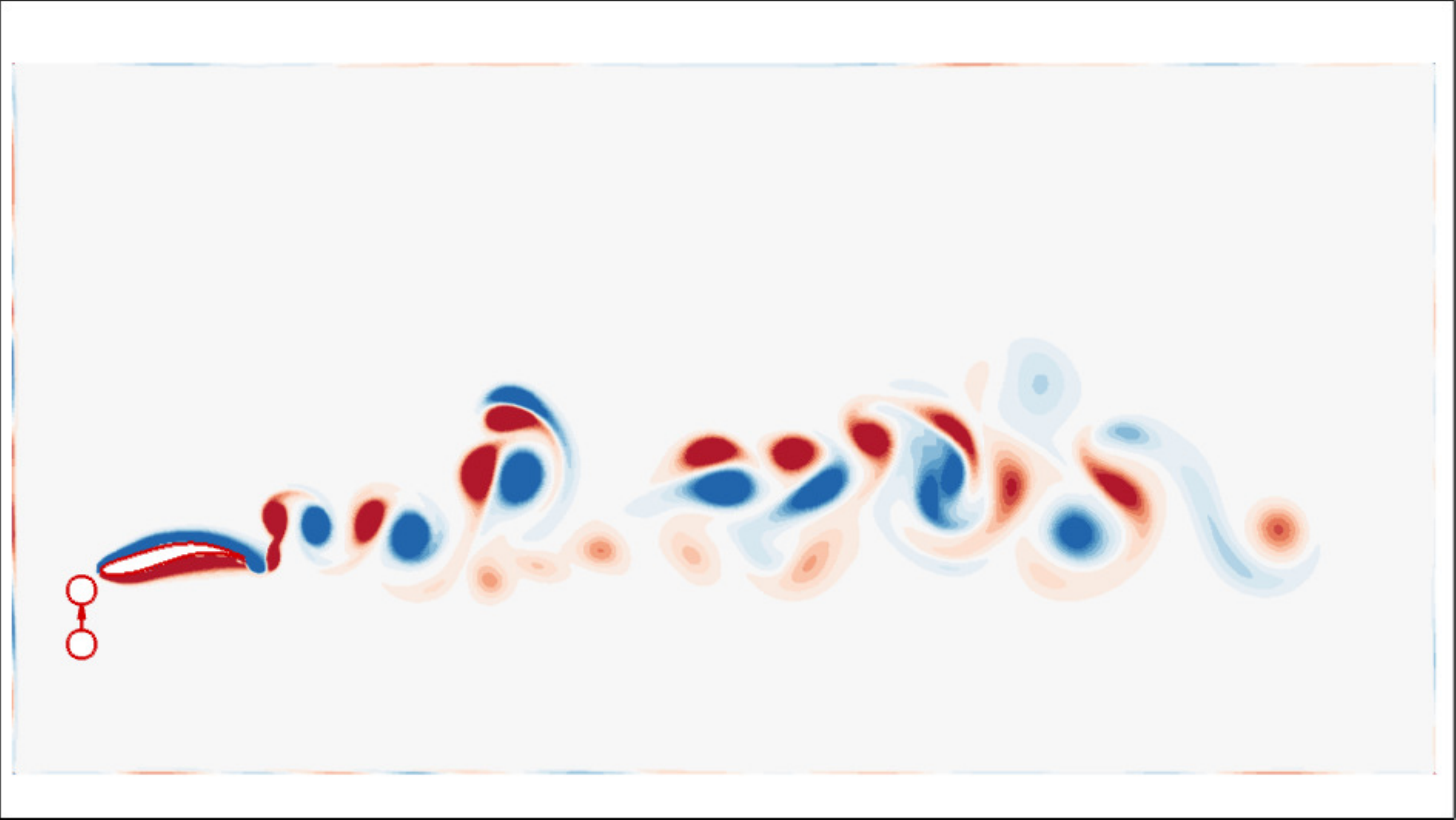}}
			\\
			\subfigure[Color bar of the vortex contour]{
				\includegraphics[width=0.48\linewidth]{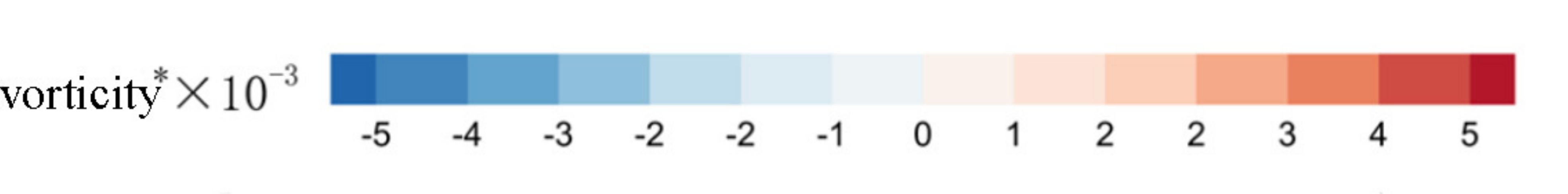}}
			\caption{Vortex contour graph of flow field in predation swimming}
			\label{fig_9_vormap1}
		\end{figure}

		Figure \ref{fig_9_vormap1} depicts the vortex contour of the flow field in this test. Before the 5.0T moment,, the food had not yet begun to move and the agent fish swam towards the food with a straight trajectory. At 6.0T, the food began to move and the fish recognized the movement signal, initiating small angle course-change maneuvers. At 11.6T, the fish continued to maneuver counterclockwise in order to follow the food's trajectory. However, at 13.3T, the fish stopped the counterclockwise maneuver and instead executed a small angle clockwise maneuver. This change in strategy was a result of the fish's ability to predict the food's trajectory and adopt a new swimming policy. By making an early clockwise maneuver, the fish was able to successfully capture the food that was moving upward. Around 14.3 waving cycles, the fish reached the food's location successfully.
		
		\subsection{Task2-K$\acute{a}$rm$\acute{a}$n gait swimming with different upstream speed}
		
		\paragraph{}
		It has been experimentally demonstrated that when the Reynolds number (Re) falls within the range of 300 to 150,000, the fluid passing through a stationary obstacle will produce a series of vortex trails, known as K$\acute{a}$rm$\acute{a}$n vortex street\cite{leonardASPECTSFLOWINDUCEDVIBRATION2001}. In nature, fish are often adept at using vortices’s energy in turbulent flows to improve their swimming efficiency and reduce the loss of biological energy. For example, when fish swim near river obstacles or when fish schools migrate, they actively extract energy from the vortices by adjusting their movement patterns with a specific waving frequency. In K$\acute{a}$rm$\acute{a}$n vortex street, fish will adopt a new swimming gait to synchronize their own body movement with the vortex shedding frequency in order to conserve energy. This unique movement pattern is called K$\acute{a}$rm$\acute{a}$n gait, which is characterized by a low waving frequency and a large lateral displacement of the body\cite{liaoRoleLateralLine2006}. The study of K$\acute{a}$rm$\acute{a}$n swimming in fish can explain the energy-saving mechanism of fish swimming and aid in the understanding of the kinetic mechanism of fish migration upstream, which has important engineering and scientific significance. Currently, computation platforms built by coupled deep reinforcement learning and fluid-structure interaction algorithms can only simulate fish swimming in turbulent flow fields at specified flow velocities\cite{vermaEfficientCollectiveSwimming,zhuNumericalStudyFish2021}. The models that have been trained cannot be reused if the flow field is changed and their generalization ability is weak. Additionally, it will take a significant amount of time to re-adapt to a new flow field if the flow field is changed and the agent fish does not have the ability to sense the flow field’s characteristic parameter like vorticity and velocity. To verify the performance of the computation platform that is combined with the lateral-line machine and macro-action system, it is proposed to use different upstream flow conditions to observe the process of the agent fish adapt to swim in different turbulence flow fields.
		
		\subsubsection{Computational parameters}
		
		1.Hyper parameters of Deep reinforcement learning module
		
		Since the action space is defined by macro-action, the state of the agent fish has to be redesigned. Considering various environmental factors ,the observed state space of the fish is designed as $[{X_{dis}},{Y_{dis}},\Delta X,\Delta Y,{U_x},{U_y},{\omega _t},{\omega _t},{\omega _t}]$, where:$X_{dis}$ is the relative distance of the fish head from the D-cylinder in the \textit{x}-direction, $Y_{dis}$ is the relative distance of the fish head from the D-cylinder in the \textit{y}-direction, $\Delta X$ is the \textit{x}-displacement of the agent fish relative to the previous moment, $\Delta Y$ is the \textit{y}-displacement of the agent fish relative to the previous moment, $U_{x}$ is the \textit{x}-direction flow velocity perceived by the lateral line of the fish, $U_{y}$ is the \textit{y}-direction flow velocity perceived by the lateral line of the fish, $\omega _t$ is the waving frequency at the current moment. In order to enable the agent fish to capture temporal information, the state information of several recent steps is stacked in chronological order as the input information for the current moment. This is because the agent fish only changes its waving frequency at every interval of one waving period. Taking into account the delay of tail data acquisition, only the state information of the last two periods is stacked. The assembled state tuple is represented in equation \ref{equ_16}. The stacked vector is then input into the deep neural network, which enables the agent fish to possess short-term tail-swing memory capability. The neural network responsible for the navigation of the fish in the flow field has been trained in Task1, and the action space of the wake is composed of the following matrix:   ${\left[ {\begin{array}{*{20}{c}}
					{\left[ {{\omega _{11}},{\omega _{12}}...{\omega _{1i}}} \right]}\\
					{\left[ {{\omega _{21}},{\omega _{22}}...{\omega _{2i}}} \right]}\\
					{...}\\
					{\left[ {{\omega _{j1}},{\omega _{j2}}...{\omega _{ji}}} \right]}
			\end{array}} \right]_{j \times i}}$, where \textit{j} is the number of flow fields in the current simulation condition, \textit{i} is the number of waving frequencies that can be selected under the current characteristic flow field, and the size of the whole solution space is $j \times i$, firstly, the current flow field type is predicted by the lateral-line machine, and the solution of the current problem is reduced to ${[{\omega _{n1}},{\omega _{n2}}...{\omega _{ni}}]_{n \in [1,j]}}$, and then the macro-action is used to find the wake frequencies that are adapted to the current turbulent flow field motion.
		
		\begin{equation}\label{equ_16}
			\left[ {\begin{array}{*{20}{c}}
					{{{[{X_{dis}},{Y_{dis}},\Delta X,\Delta Y,{U_x},{U_y},{\omega _t},{\omega _t},{\omega _t}]}_{\rm{n}}}}\\
					{{{[{X_{dis}},{Y_{dis}},\Delta X,\Delta Y,{U_x},{U_y},{\omega _{t - 1}},{\omega _{t - 1}},{\omega _{t - 1}}]}_{n - 1}}}
			\end{array}} \right]
		\end{equation}
	
		The design of action space, state space and reward function are closely interrelated and are designed in a comprehensive manner to enable the agent fish to complete the task effectively. To take into account the position and energy consumption of the agent fish, the total reward is composed of two main parts, which are the primary reward and the secondary reward. For the primary reward: 1. If the fish is in the current flow field $\Delta X > 20$,illustrate that the current tail frequency cannot sustain the stable swimming of the fish in the turbulent flow field, $r_{1}=-1$ , otherwise $r_{1}=1$; 2. If the fish swims to the dangerous area in the flow field, $r_{3}=-5$ ,otherwise $r_{3}=0$; 3. $r_{4}$ is the potential energy, If the fish is approaching the dangerous swimming area, ${{\rm{r}}_4} =  - ({L_{\max }} - {L_{dis}})/10$ , where ${L_{max }}$ is the maximum distance the fish is allowed to swim from the dangerous area, and ${L_{dis }}$ is the current distance of the fish from the dangerous area. For the secondary reward $r_{2}$:${r_2} = 0.1*(1 - {w_{period}})$,where ${w_{period}}$ is the normalized work in a waving period, and the work is higher, the reward is lower, which is introduced to take into account the energy consumption of the fish swimming.
		
		The hyperparameter settings for training are shown in table \ref{tab1}, and the policy loss entropy temperature   is performed by adaptive computation\cite{haarnojaSoftActorCriticAlgorithms2019}.
		
		\begin{table}[]
			\centering
			\small
			\caption{Hyper-parameter of DRL}
			\label{tab1}
			\begin{tabular}{cm{15mm}<{\centering}m{15mm}<{\centering}m{15mm}<{\centering}m{15mm}<{\centering}m{15mm}<{\centering}m{15mm}<{\centering}}
				\hline
				& Learning rate $\alpha$ & Discount factor$\gamma$ & Buffer size $N_{D}$ & Soft update parameters $\tau$ &
				Batch size $N_{B}$ & Environment step$n$
				 \\ \hline
				Parameter& 0.0005 & 0.99 & 500000 & 0.01 & 128 & 100T \\
			
			 \hline
			\end{tabular}
		\end{table}
		
		2.Hyper parameters of fluid solver
		
		For the fluid solver module, the characteristic parameters of the flow field are the same as in the previous section, and the characteristic waving period of the fish body is still taken as the characteristic time of the flow field, where the characteristic waving period is 1000 dimontionless lattice time, and the boundary conditions are as follows: the velocity inlet boundary on the left, the free outflow boundary on the right, the top and bottom are non-equilibrium extrapolation boundary, the density ${\rho ^*}$  of the flow field is set to 1 and the diameter of the D-cylinder used to generate the vortex is 0.4L, and the upstream flow velocity is 1.00L/T, 1.25L/T, 1.50L/T, where L is the length of the fish, and the Reynolds number at different flow rates are 200, 250, and 300, respectively. This case has been implemented on a workstation equipped with an Nvidia RTX A4000 graphics card, utilizing Nvidia's CUDA technology for GPU parallel computing.
		

		\begin{table}[]
		\centering
		\small
		\caption{Flow field parameter setting}
		\label{tab2}
			\begin{tabular}{cm{20mm}<{\centering}m{20mm}<{\centering}m{20mm}<{\centering}m{20mm}<{\centering}}
				\hline
				& Upstream flow velocity $u^{*}$ & Relaxation time $\tau$ & Renolds number $Re$ & Vortex shedding frequecy $f^{*}*10^{-4}$  \\ \hline
			Working condition \RNum{1}	& 1.00 & 0.56 & 200 & 6.67 \\
			Working condition \RNum{2}& 1.25 & 0.56 & 250 & 8.00 \\
			Working condition \RNum{3}& 1.50 & 0.56 & 300 & 10.00 \\ \hline
			\end{tabular}
		\end{table}
		
		\subsubsection{Numerical simulation results}
		
		\paragraph{}
		In order to verify the reliability of the agent fish with macro-action system and lateral-line machine and to demonstrate the efficiency of its knowledge transfer system, the swimming decision system of the agent fish was first trained to convergence under working condition II, in which the upstream speed $u^{*}$ is set as 1.25L/T. The trained fish was then placed into the turbulent flow field of working condition I and working condition III to observe whether it could adapt to the swimming task of the complex flow field by utilizing lateral line perception.

\begin{figure}[!t]
			\centering
			\includegraphics[width=5in]{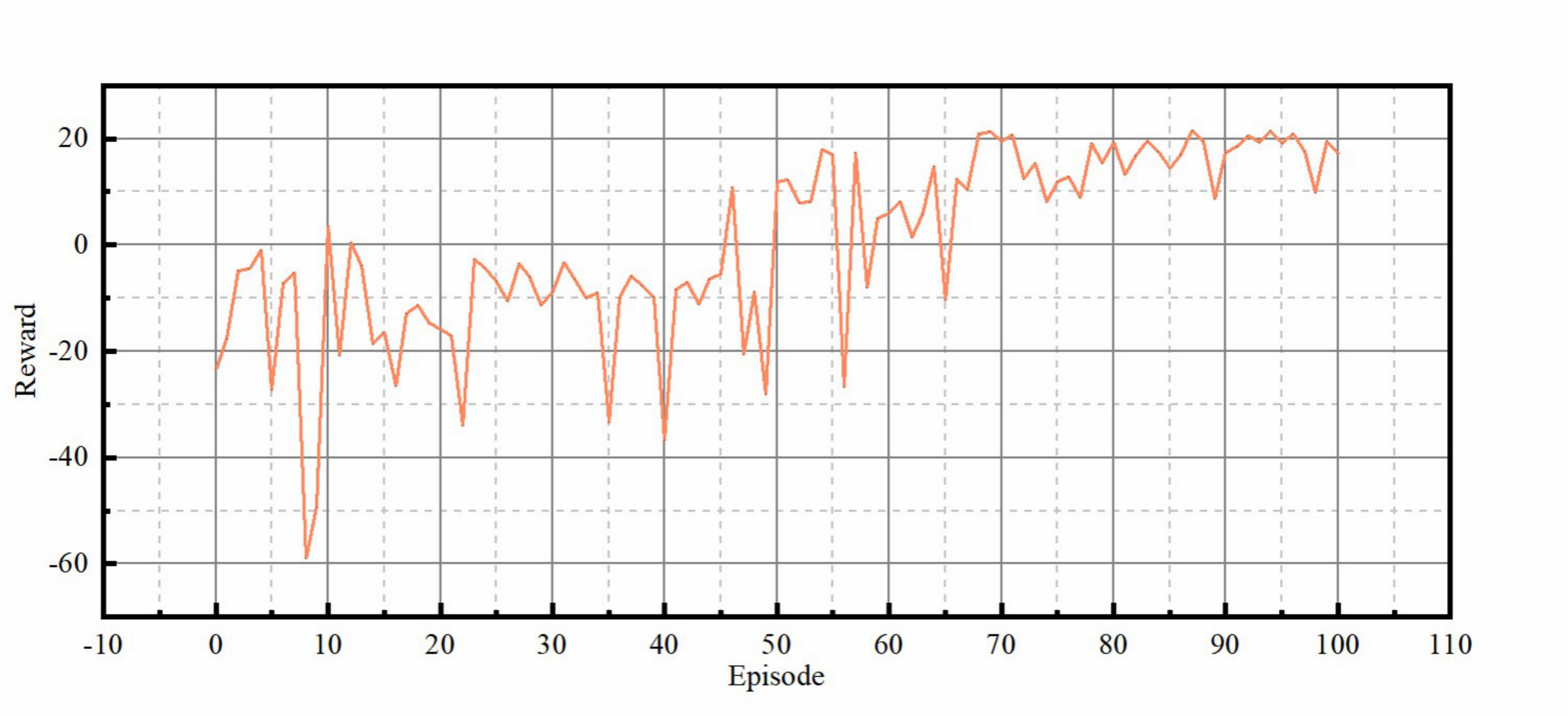}
			\caption{Reward received by the agent fish in the K$\acute{a}$rm$\acute{a}$n gait test}
			\label{fig_10_reward2}
		\end{figure}

\begin{figure}[!t]
			\centering
			\includegraphics[width=5in]{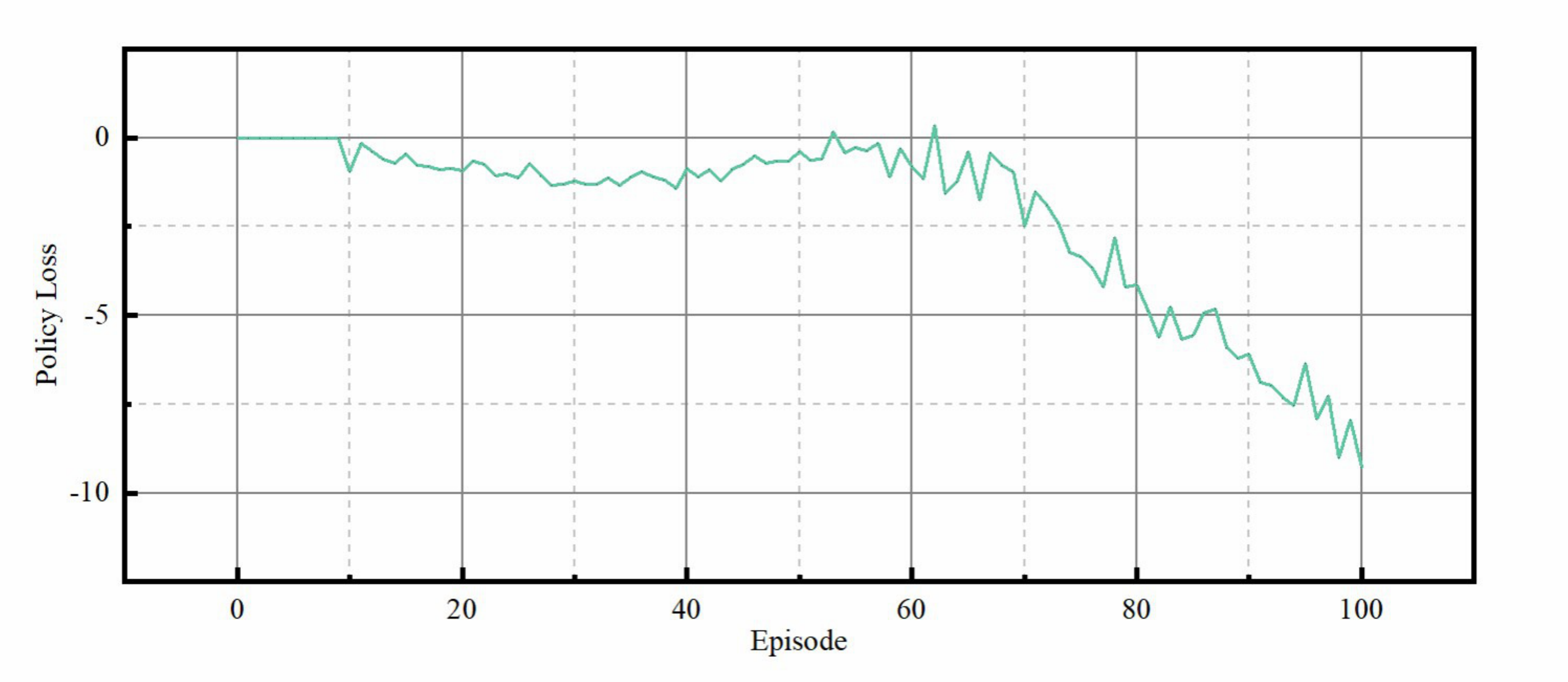}
			\caption{Policy loss of the actor policy network in the K$\acute{a}$rm$\acute{a}$n gait test}
			\label{fig_11_loss2}
		\end{figure}

		Figure \ref{fig_10_reward2} shows the reward received by the fish in each episode of training, and figure \ref{fig_11_loss2} shows the loss of the actor policy network in the K$\acute{a}$rm$\acute{a}$n gait test. Due to the complexity and high nonlinearity of the turbulent flow field environment, the reward is not stable and appears to fluctuate for a prolonged period until the 70th episode, indicating that the agent fish has not yet found an optimal swimming policy for the complex flow field at this stage. However, after 70 episodes, the reward reaches a high and stable level, and the policy loss gradually decreases, indicating that the agent fish has learned the correct swimming policy through previous training, and the training results tend to converge. As a result, the fish that has undergone stable training in working condition II can be directly placed in the flow field of working condition I and working condition III to observe its swimming behavior in unfamiliar environments and verify the knowledge transition system.\\
		
		\textbf{(1)New development module's advantages}
		
		To illustrate the advantages of the agent fish with lateral-line machine and macro-action system, the same deep reinforcement learning algorithm was used, however,but the lateral-line machine and macro-action system were removed, and a single waving frequency action space $\left[ {{\omega _1},{\omega _2}...{\omega _n}} \right]$ was implemented. For example, the agent can only select one action  $\omega_{i}$ in $\left[ {{\omega _1},{\omega _2}...{\omega _n}} \right]$ at time t. This was done to compare the ability of the fish to maintain its position in different turbulent flow fields. Firstly, two different agent fish were trained for the same time and placed under three different flow fields. Each test was repeated 100 episodes, and the agent's survival time in each episode in the turbulent flow field was recorded. The upper limit of swimming time for each episode is 100T.

\begin{figure}[H]
			\centering  
			\subfigbottomskip=2pt 
			\subfigcapskip=-5pt
			\subfigure[Survival time of fish in the turbulent flow field when the lateral-line machine and macro-action system is banned(The experiment was repeated a total of 100 times.)]{
				\includegraphics[width=0.48\linewidth]{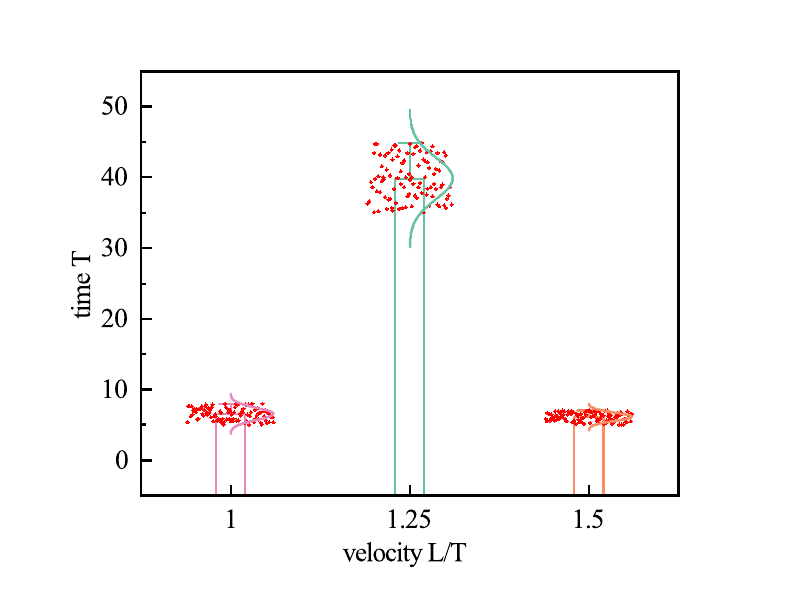}}
			\subfigure[Survival time of agent fish in turbulent flow when the lateral-line machine and macro-action system is activated(The experiment was repeated a total of 100 times.)]{
				\includegraphics[width=0.48\linewidth]{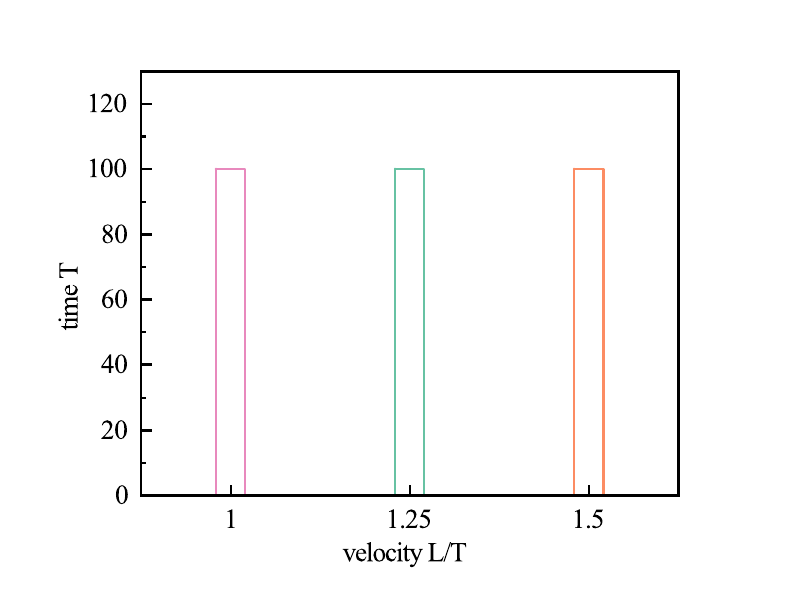}}
			\caption{Graph of comparison of survival time}
			\label{fig_12_swimtime}
		\end{figure}

		Figure \ref{fig_12_swimtime} (a) illustrates the survival time of the fish in the turbulent flow field when the lateral-line machine and macro-action system is banned. It can be observed that the average survival time of the agent fish in three different turbulent flow fields is relatively short, with average survival times of 6.54T, 39.80T, and 6.13T respectively. The reason for the relatively high survival time of 39.80T in working condition II is that the agent fish was first trained in working condition II, making it more adaptable to the flow field of working condition II. However, since the agent fish does not have the lateral-line machine and macro-action system, its poor generalization ability is demonstrated in this test. When the fish encounters different turbulent flow fields, the knowledge learned in other flow fields cannot be transferred. Thus, after changing the flow field environment, the trained swimming policy fails and the survival time is at a low level. Figure \ref{fig_12_swimtime} (b) illustrates the survival time of the agent fish in turbulent flow when the lateral-line machine and macro-action system are activated. It can be observed that the agent fish has a significantly robust ability to maintain its position in turbulent flow, and the stability of its movement is also significantly enhanced. The survival time of the fish in all episodes of all working conditions reaches the maximum episode time of 100T, which fully illustrates the advantages of using the lateral-line machine and macro-action system.
		
\begin{figure}[H]
			\centering  
			\subfigbottomskip=2pt
			\subfigcapskip=-5pt 
			\subfigure[Trajectory of the agent fish without the lateral-line machine and macro-action system]{
				\includegraphics[width=0.48\linewidth]{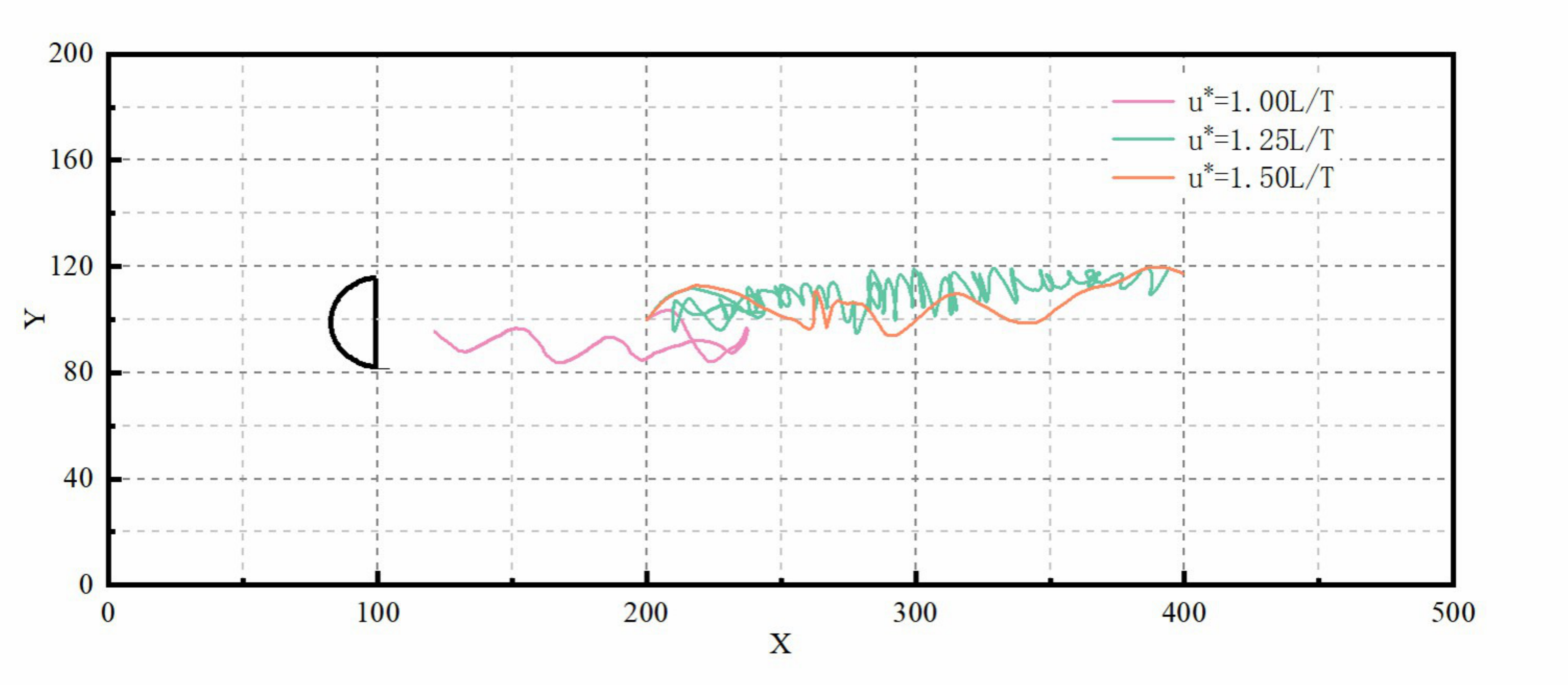}}
			\subfigure[Trajectory of the agent fish using the lateral-line machine and the macro-action system]{
				\includegraphics[width=0.48\linewidth]{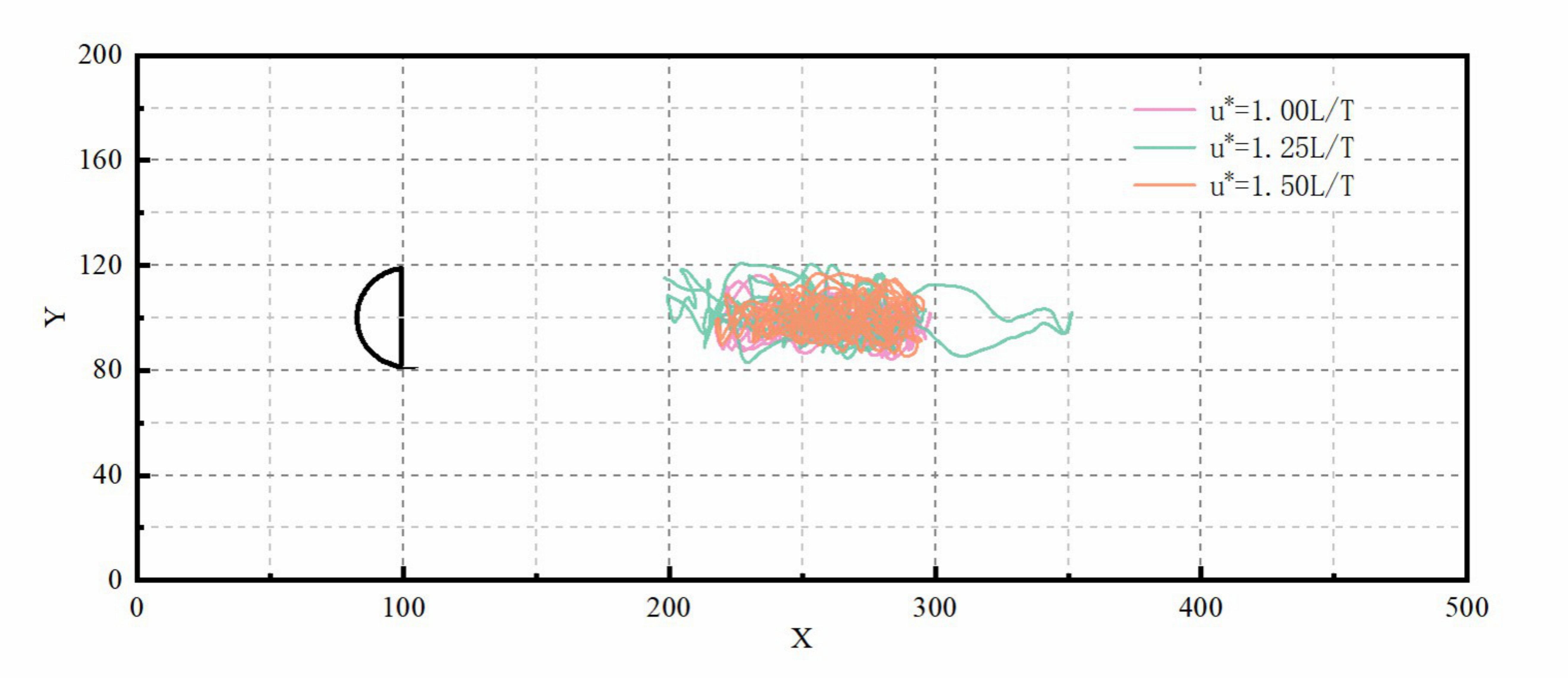}}
			\\
			\subfigure[Experimental results\cite{liaoRoleLateralLine2006}]{
				\includegraphics[width=0.48\linewidth]{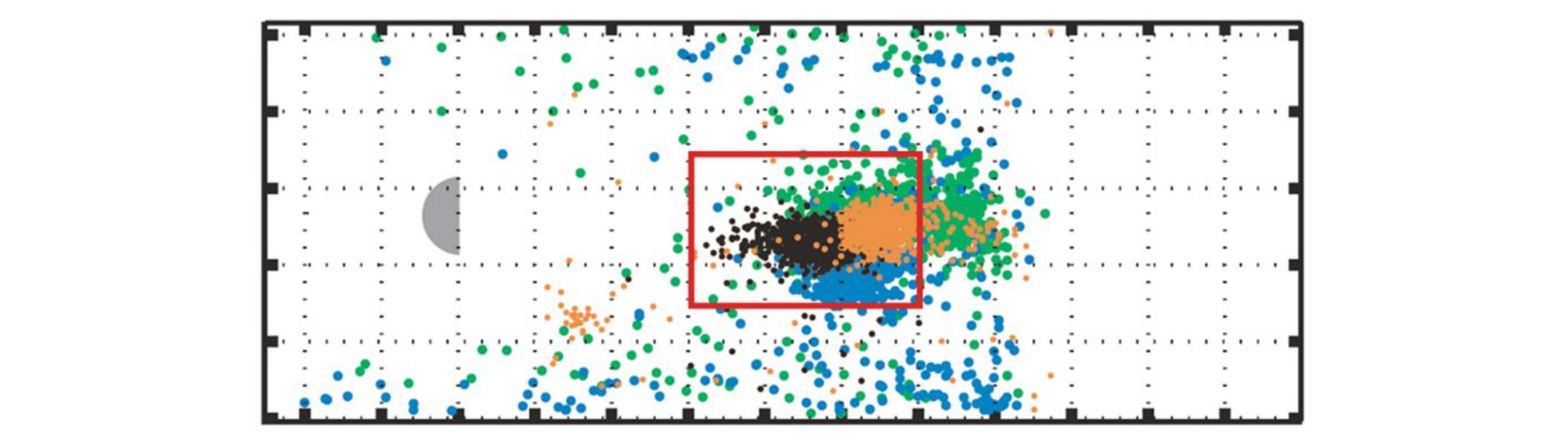}}
			\caption{Graph of comparison of fish's trajectory in turbulent field}
			\label{fig_13_locus}
		\end{figure}

		Figure \ref{fig_13_locus} (a) illustrates the trajectory of the fish without the lateral-line machine and macro-action system. Since the fish does not have the ability to recognize the flow field and cannot take macro-action maneuvers such as various accelerations and decelerations according to its own learning, it is forced to adopt a cruising mode with a limited number of waving frequencies. The performance of this model is poor in turbulent flow, and the maximum survival time of the agent fish using this swimming policy is 39.8T.Figure \ref{fig_13_locus} (b) shows the trajectory of the fish using the lateral-line machine and the macro-action system. The fish adopted a similar swimming decision in the three different turbulent flow fields. The swimming policy in the K$\acute{a}$rm$\acute{a}$n vortex street is analyzed as follows: The agent primarily adopts the K$\acute{a}$rm$\acute{a}$n gait and uses acceleration and deceleration maneuvers as auxiliary. In terms of swimming position, the agent fish chooses to swim in the area between the two vortex streets in order to save swimming energy. The \textit{Y}-direction displacement of the agent fish does not exceed the diameter range of the D-cylinder, so as to extract energy from the alternating vortices. While in the position parallel to the flow direction, the fish body chooses a position 1-2 body lengths away from the D-cylinder using the K$\acute{a}$rm$\acute{a}$n gait, which is in good agreement with previous experimental results\cite{liaoRoleLateralLine2006}. The survival time of the agent fish in all working conditions reached the maximum episode time of 100T. This suggests that the swimming policy of the agent fish with the lateral-line machine is more similar to that of live fish and can make different complex maneuvers to adapt to the turbulent flow. The use of the lateral-line machine and macro-action system enables the agent fish to better recognize and respond to the flow field, thus improving its performance and survival time in different turbulent flow conditions.
		
\begin{figure}[H]
			\centering  
			\subfigbottomskip=2pt 
			\subfigcapskip=-5pt 
			\subfigure[$t^{*}=4.0T$]{
				\includegraphics[width=0.48\linewidth]{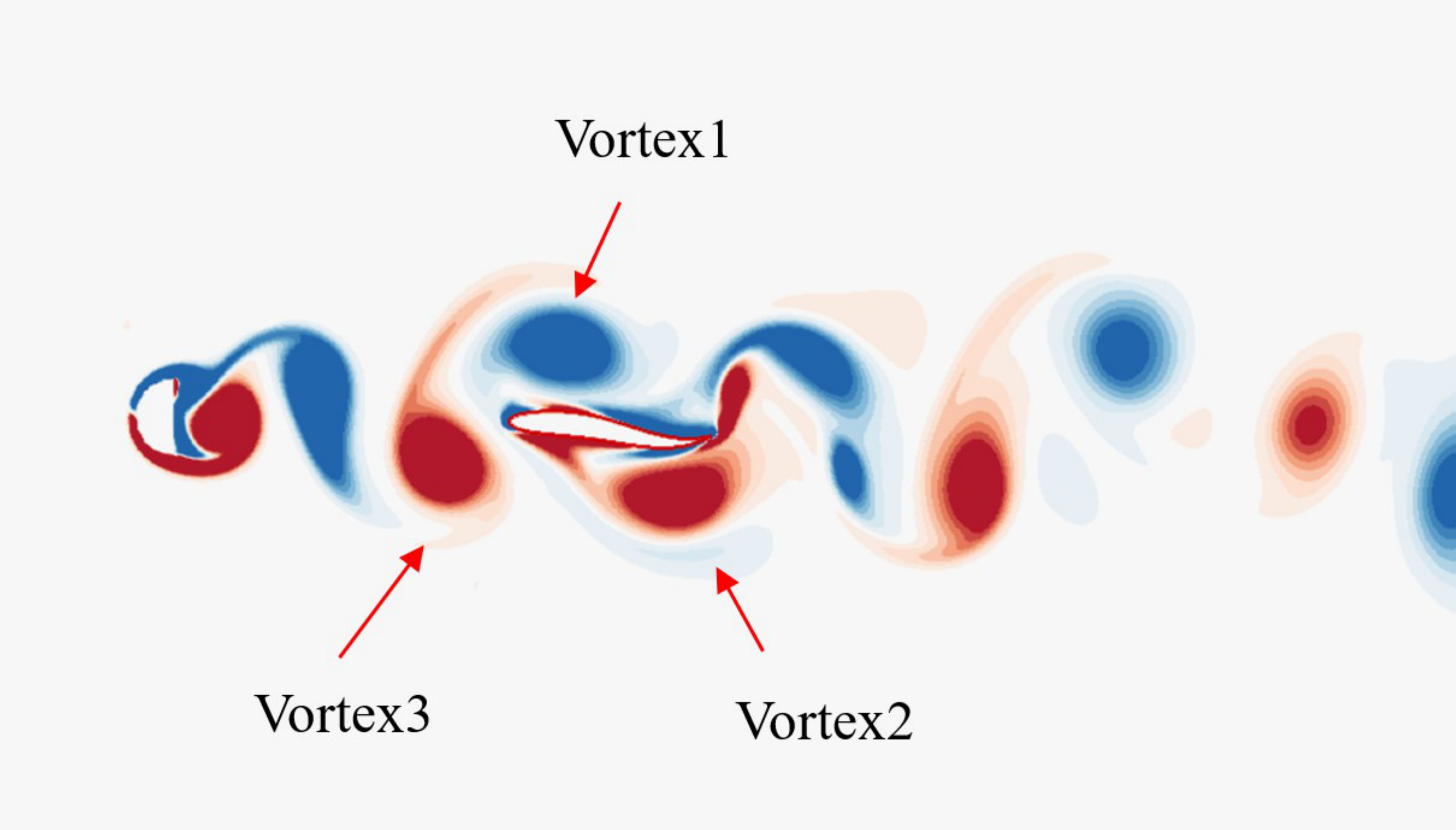}}
			\subfigure[$t^{*}=4.3T$]{
				\includegraphics[width=0.48\linewidth]{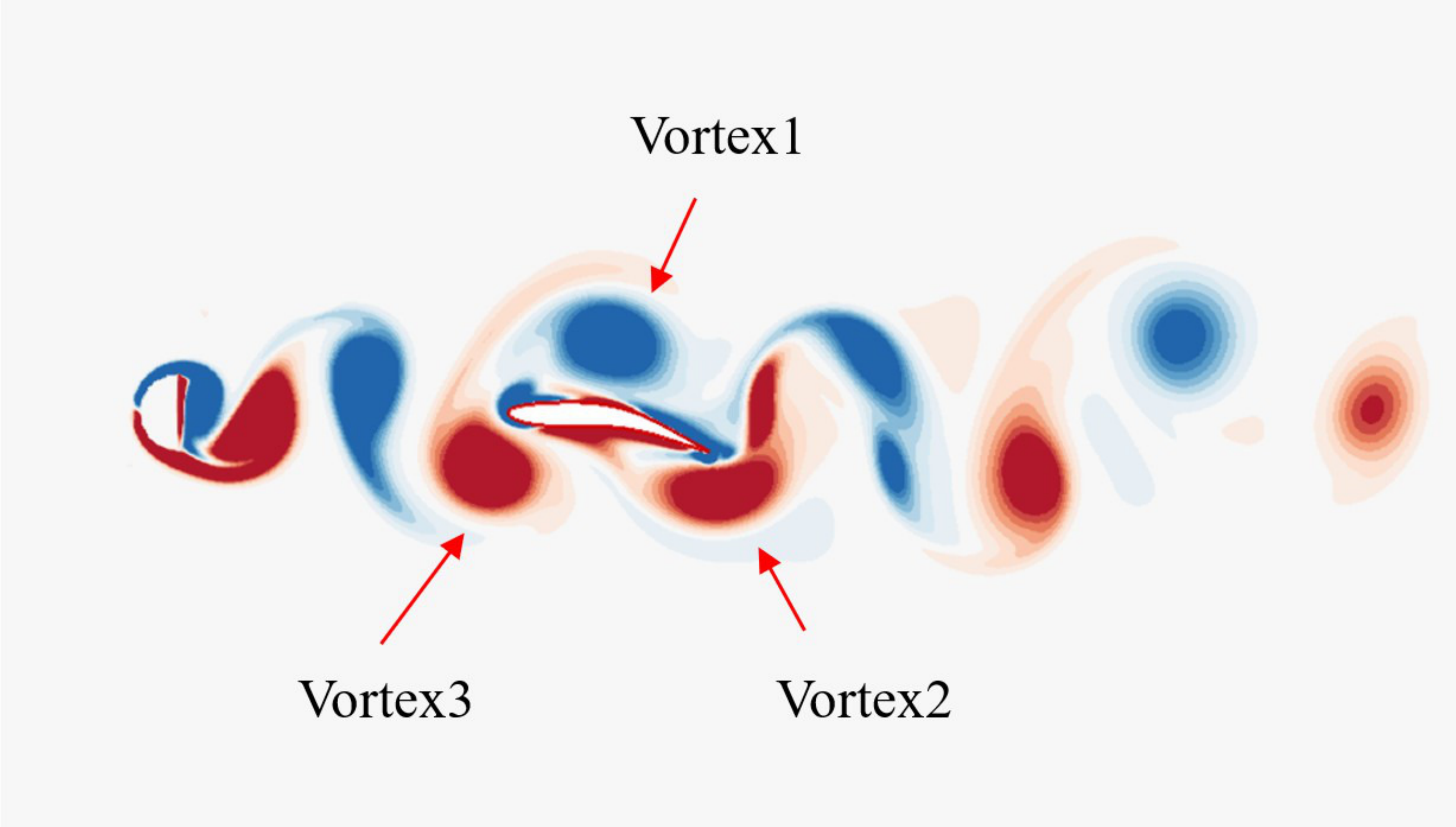}}
			\\
			\subfigure[$t^{*}=4.6T$]{
				\includegraphics[width=0.48\linewidth]{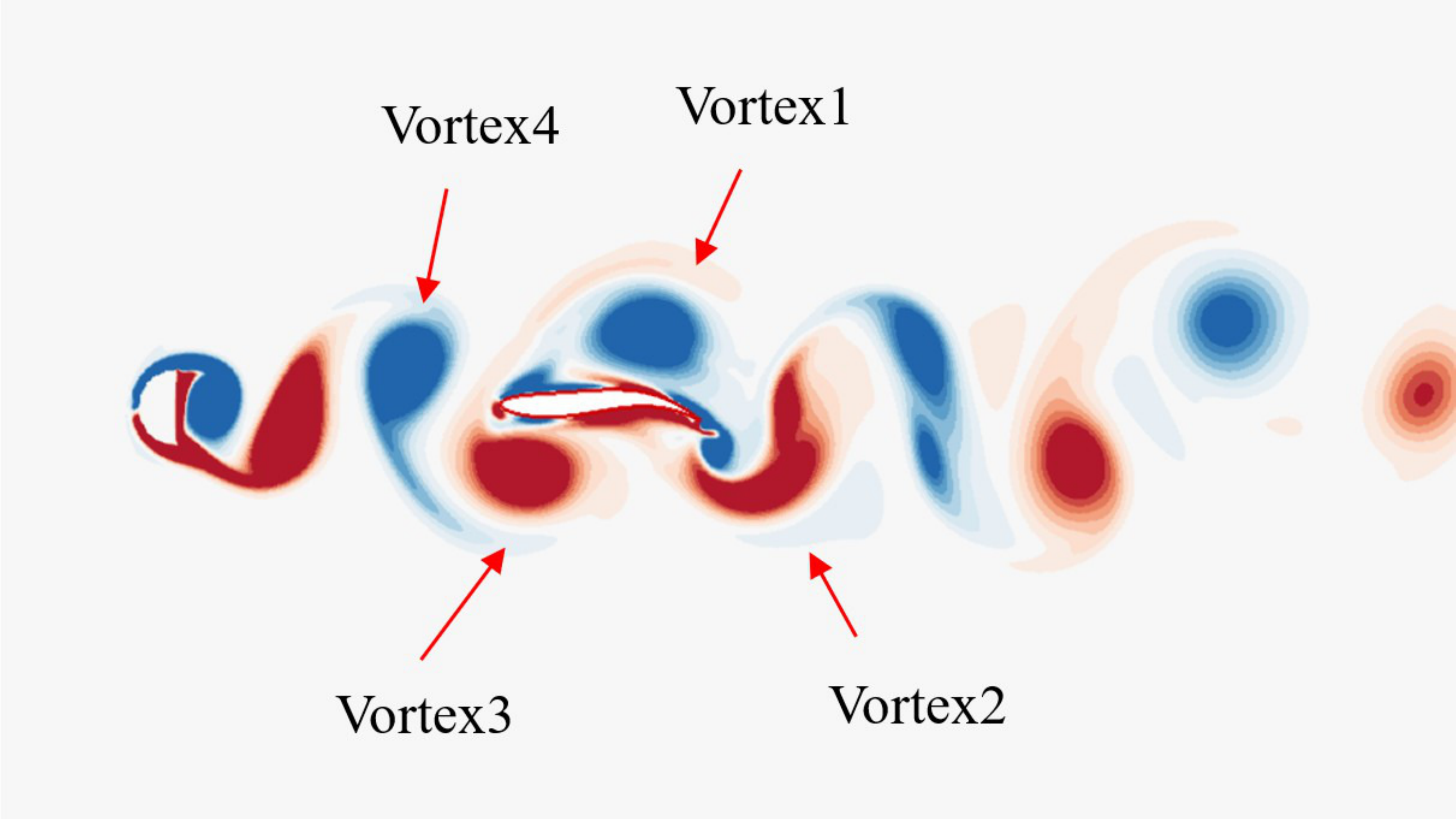}}
			\subfigure[$t^{*}=4.9T$]{
				\includegraphics[width=0.48\linewidth]{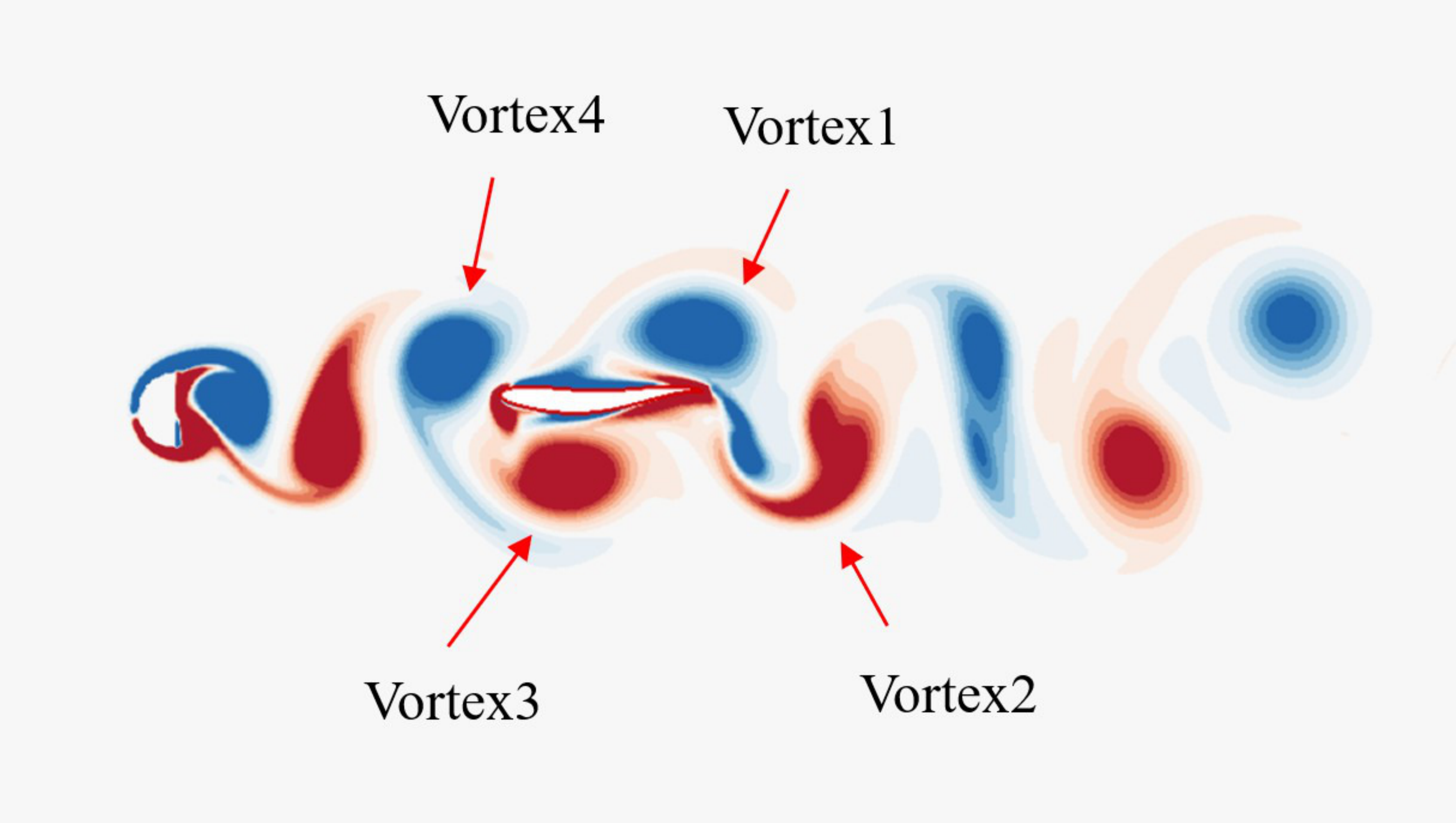}}
			\\
			\subfigure[$t^{*}=5.2T$]{
				\includegraphics[width=0.48\linewidth]{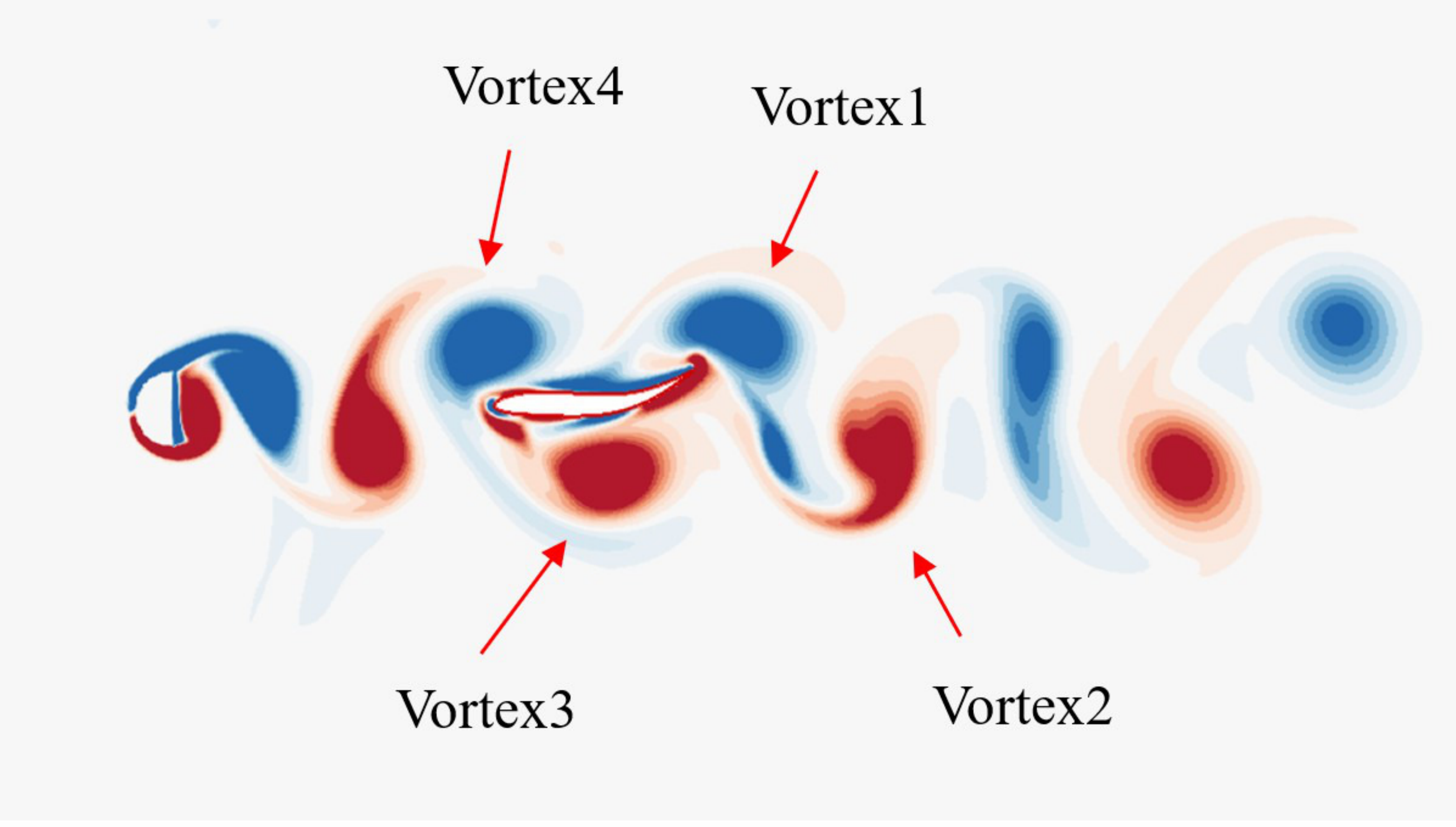}}
			\subfigure[$t^{*}=5.5T$]{
				\includegraphics[width=0.48\linewidth]{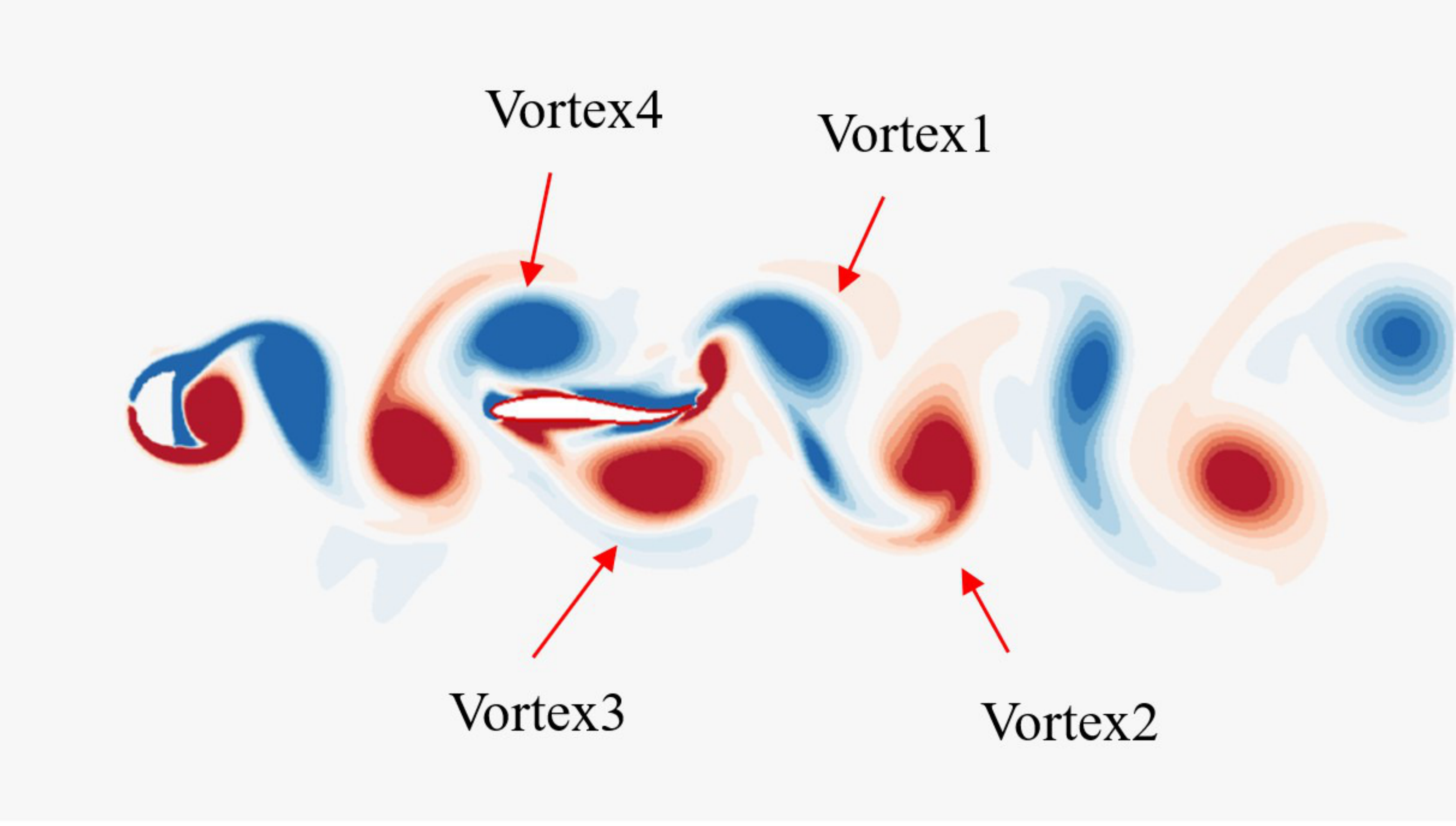}}
			\\
			\subfigure[Color bar of the vortex contour]{
				\includegraphics[width=0.48\linewidth]{color_bar-eps-converted-to.pdf}}
			\caption{Vortex contour graph of the agent fish with a lateral-line machine and macro-action system in a waving period in the K$\acute{a}$rm$\acute{a}$n gait test}
			\label{fig_14_vormap2}
		\end{figure}

		Figure \ref{fig_14_vormap2} illustrates the vortex contour map of the agent fish with a lateral-line machine in a waving period,where ${u^*}{\rm{ = }}1.00{\mathop{\rm L}\nolimits} /T$. It can be clearly observed that the agent fish extracts energy from the upper and lower vortices. At 4.0T, the upper side of the agent fish impacts a clockwise vortex (vortex1) while the counter-clockwise vortex (vortex2) has slid from the head to the tail of the agent fish. At 4.6T, vortex1 slides from the head to the tail of the agent fish and vortex2 separates from the fish's tail, followed by the head part of the agent fish impacting the counter-clockwise vortex3. At 5.5T, vortex1 separates from the fish's tail and the fish's head impacts the clockwise vortex4, vortex3 slides from the head to the tail of the agent fish. This figure clearly demonstrates the fish's K$\acute{a}$rm$\acute{a}$n gait in turbulent flow and qualitatively explains how the agent fish extracts energy from the vortex street. During the swimming process, the vortices alternately form high shear-stress regions on the upper and lower sides of the agent fish, which helps it swim more effectively.\\
		\textbf{(2)Analysis of agent fish's swimming strategy in  K$\acute{a}$rm$\acute{a}$n vortex street}
		
		Table \ref{tab3} presents the swimming characteristics of the agent fish with the lateral-line machine. From the decision of the waving frequency of the fish body, the agent fish actively chooses a waving frequency that is consistent with the vortex street shedding frequency, allowing the fish to extract energy from the vortex and maintain its position in the turbulent flow. In working condition 1, the agent fish made a total of 5 accelerations, accounting for 4.5\% of the total time, and the remaining 95.5\% of the time was spent performing K$\acute{a}$rm$\acute{a}$n gait with a waving frequency similar to the vortex shedding frequency. In working condition 2, the agent fish made a total of 10 accelerations and 2 decelerations, accounting for 8.0\% and 5.0\% of the total time, respectively, and the K$\acute{a}$rm$\acute{a}$n gait time was 87.0\%. In working condition 3, the agent fish made a total of 10 acceleration movements, accounting for 7.0\% of the total time, and K$\acute{a}$rm$\acute{a}$n gait time was 93.0\%. This indicates that the fish spends a significant amount of time in the high-energy-efficient K$\acute{a}$rm$\acute{a}$n gait and only a small amount of time in acceleration and deceleration maneuvers to maintain its position in the turbulent flow field.
		
			\begin{table}[]
			\centering
			\small
			\caption{Flow field parameter setting}
			\label{tab3}
			\begin{tabular}{cm{20mm}<{\centering}m{20mm}<{\centering}m{20mm}<{\centering}}
				\hline
				& Vortex shedding frequency & Tail waving frequency & Proportion of K$\acute{a}$m$\acute{a}$n gaiting(\%)  \\ \hline
				Working condition \RNum{1}	& 0.00067 & 0.00067±0.0001 & 95.5  \\
				Working condition \RNum{2}& 0.0008 & 0.0008±0.0001 & 87.0  \\
				Working condition \RNum{3}& 0.001 & 0.001±0.0001 & 93.0  \\ \hline
			\end{tabular}
		\end{table}
		
\begin{figure}[!t]
			\centering
			\includegraphics[width=5in]{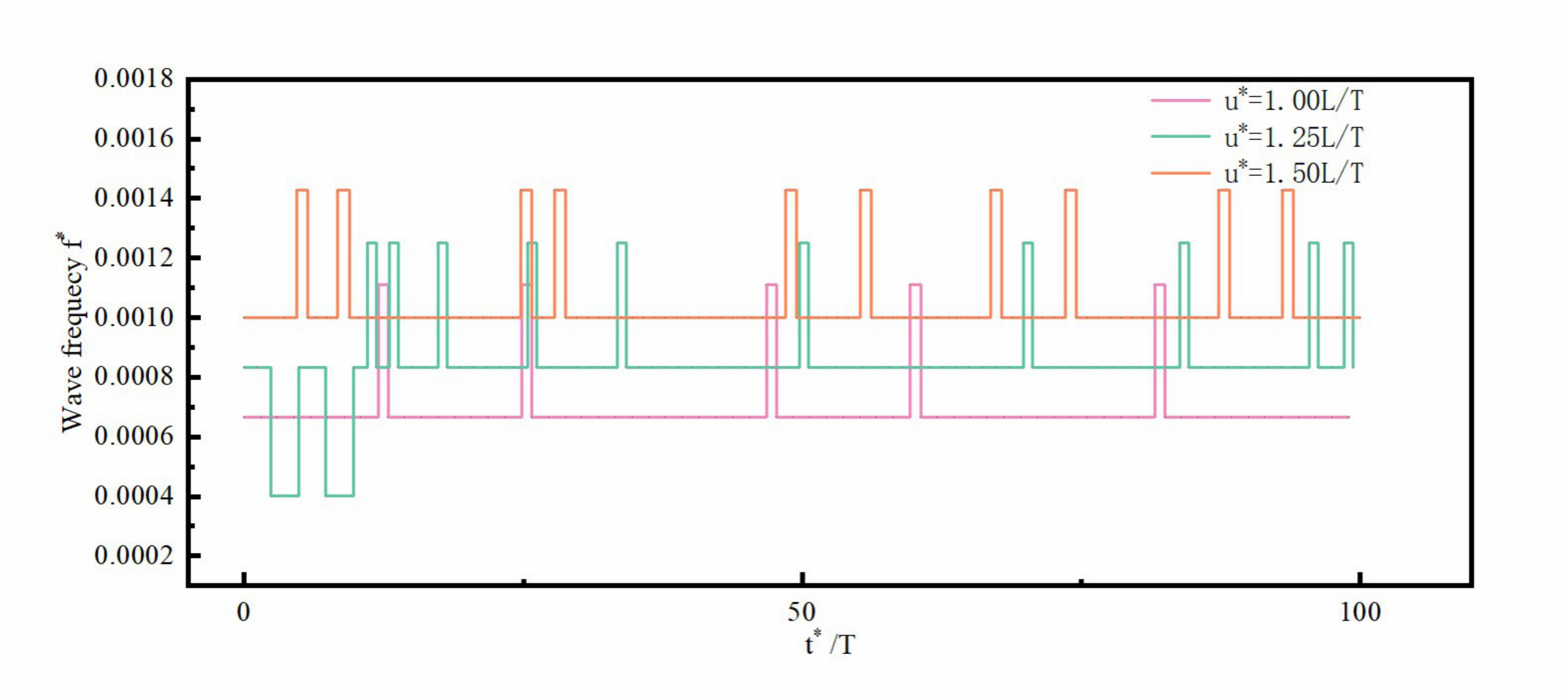}
			\caption{Time history graph of the agent fish's decision-making for waving frequency}
			\label{fig_15_wavingfeq}
		\end{figure}

		Figure \ref{fig_15_wavingfeq} shows the agent fish's decision-making for waving frequency in different turbulent flow fields. It can be observed that when navigating in different complex flow fields, the agent fish perceives different turbulent flow information to choose different swimming policies to maintain its position in the turbulent flow field, which is primarily composed of K$\acute{a}$rm$\acute{a}$n gait, supplemented by acceleration and deceleration maneuvers. The advantage of using complex maneuvers is that the fish can dynamically adjust its maneuvering policy according to its swimming state in real-time, enabling trained knowledge transfer in different flow fields, and avoiding being swept downstream by the vortex. The agent fish with a lateral-line machine and macro-action system exhibits excellent adaptability to different turbulent flow fields, and its decision system is robust and possesses strong generalization ability.
		
\begin{figure}[!t]
			\centering
			\includegraphics[width=5in]{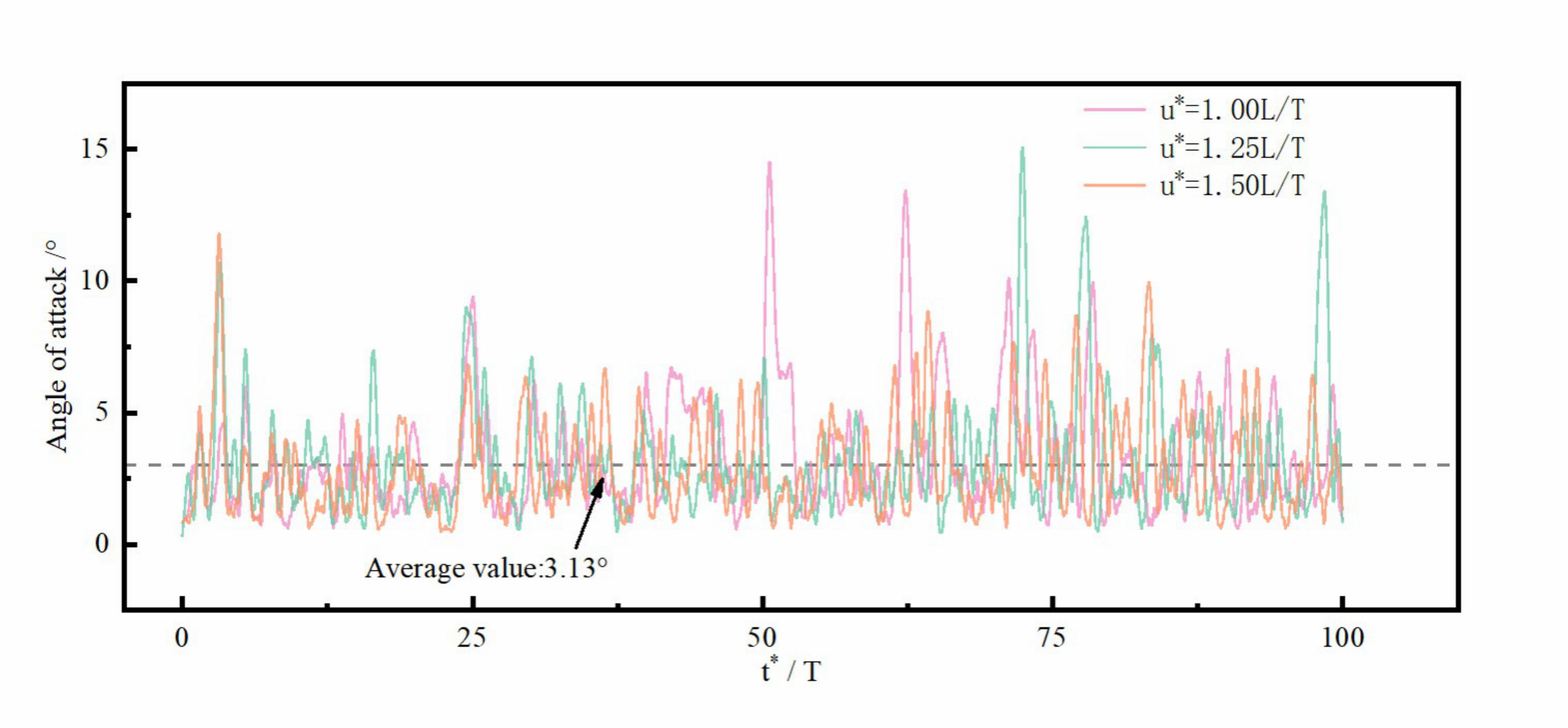}
			\caption{Time history graph of the angle of attack of the agent fish swimming in different flow fields}
			\label{fig_16_aoa}
		\end{figure}

		Figure \ref{fig_16_aoa} shows the time history of the angle of attack of the agent fish swimming in different flow fields. The angle of attack is defined as the angle of deviation of the fish body relative to the upstream flow direction. The mean value of the angle of attack for multiple conditions in this study is 3.13°. The maximum angle of attack for working condition \RNum{1} is 14.3°, for working condition \RNum{2} is 15.0°, and for working condition \RNum{3} is 11.7°. These numerical results are in close agreement with the experimental results reported by Liao\cite{liaoRoleLateralLine2006}.
		
\begin{figure}[!t]
			\centering
			\includegraphics[width=5in]{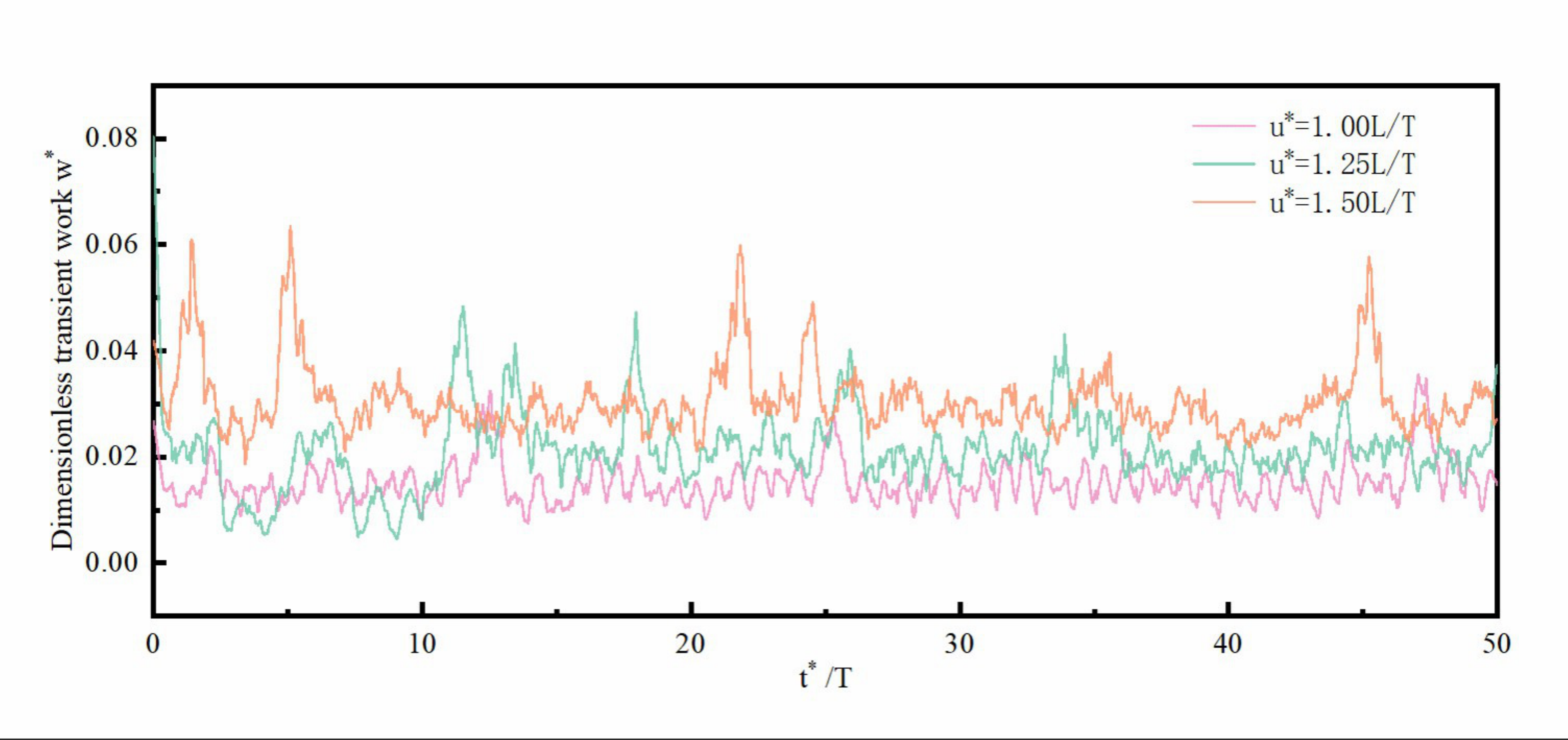}
			\caption{Graph of the transient work of the K$\acute{a}$m$\acute{a}$n gait during the first 50T}
			\label{fig_17_workall}
		\end{figure}

		As depicted in figure \ref{fig_17_workall}, the transient work of the K$\acute{a}$m$\acute{a}$n gait during the first 50T is illustrated under three different working conditions. It can be observed that the transient work under these conditions exhibits a regular fluctuation trend. The reason for the wave peaks and troughs in the work history can be attributed to the acceleration or deceleration maneuvers taken by the agent fish in a turbulent field. The stable waves, on the other hand, correspond to the K$\acute{a}$m$\acute{a}$n gait. The waves resulting from acceleration maneuvers have shorter wavelengths and higher peaks, while those from deceleration maneuvers have longer wavelengths and lower troughs. In contrast, the waves from the K$\acute{a}$m$\acute{a}$n gait remain in a stable, small-amplitude oscillation state.
		
\begin{figure}[H]
			\centering  
			\subfigbottomskip=2pt 
			\subfigcapskip=-5pt 
			\subfigure[$t^{*}=4.0T$]{
				\includegraphics[width=0.48\linewidth]{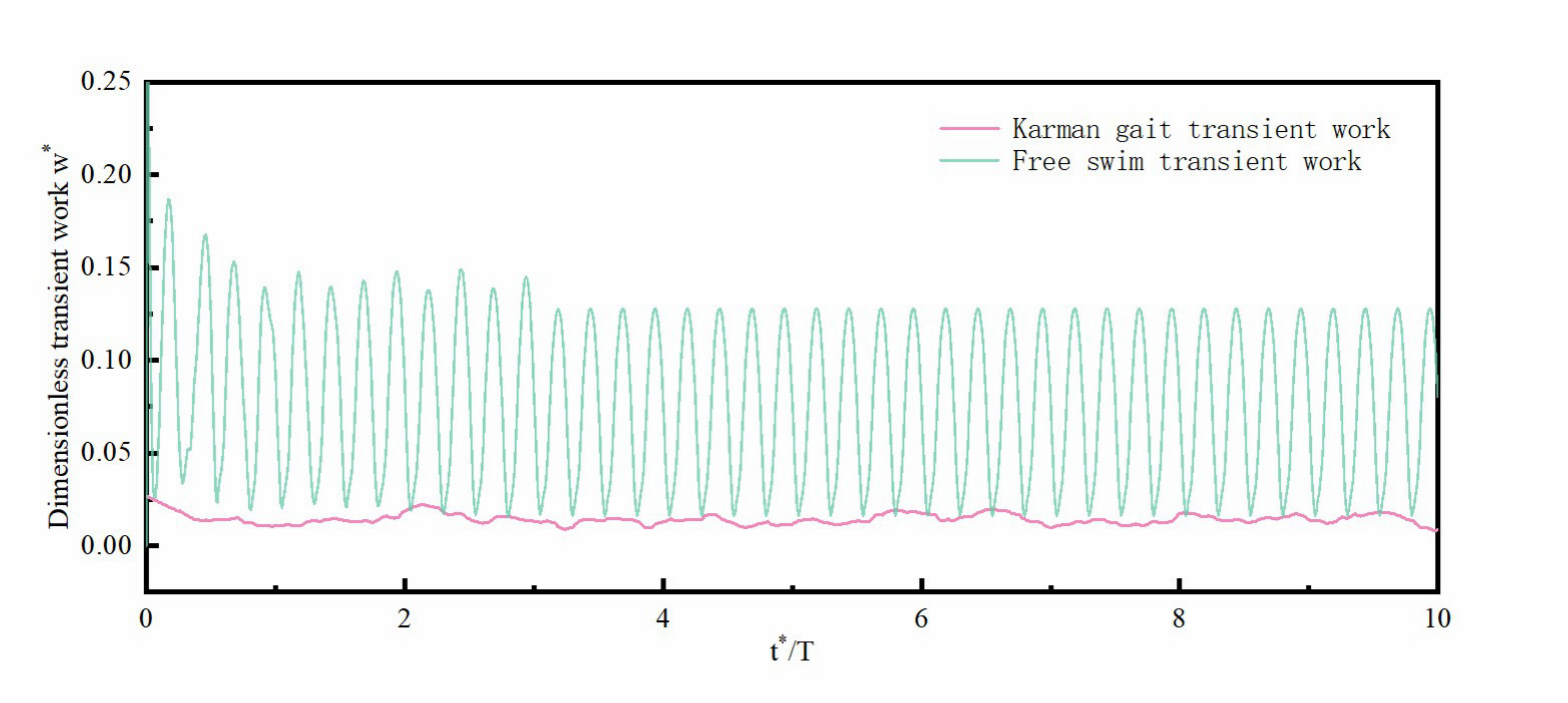}}
			\subfigure[$t^{*}=4.3T$]{
				\includegraphics[width=0.48\linewidth]{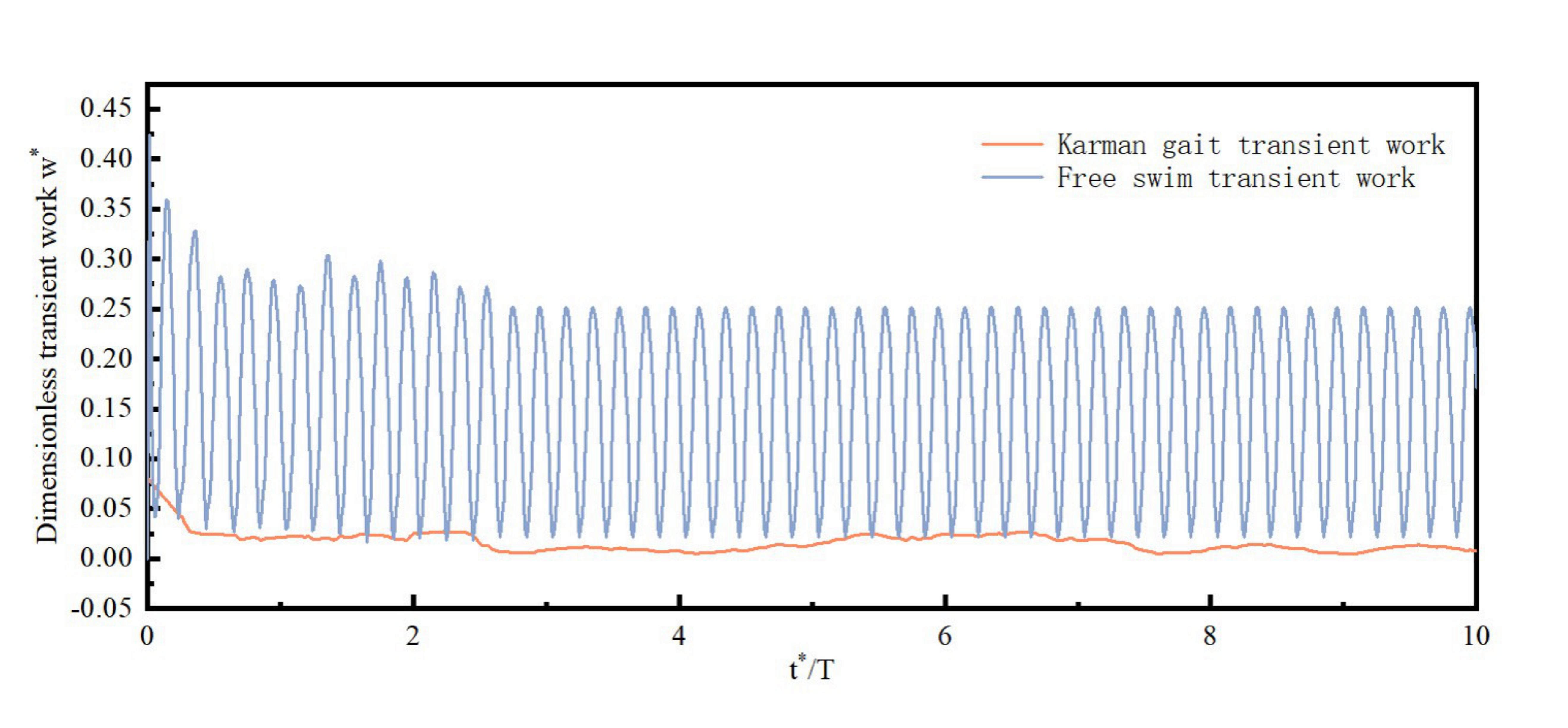}}
			\\
			\subfigure[$t^{*}=4.6T$]{
				\includegraphics[width=0.48\linewidth]{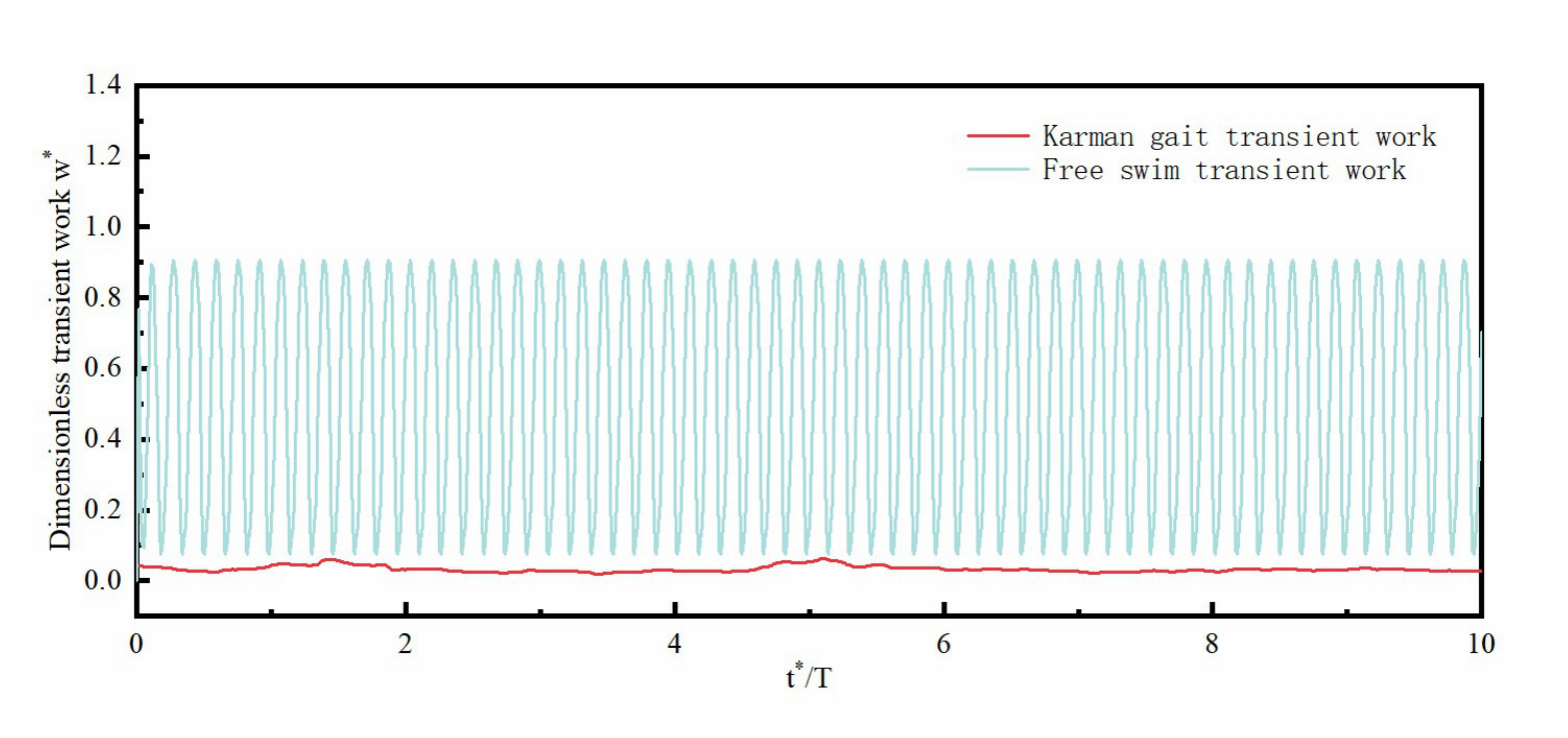}}
			\caption{Graph of the transient work time history of the fish in both the K$\acute{a}$m$\acute{a}$n vortex and free-swimming conditions for three different upstream speeds}
			\label{fig_18_work}
		\end{figure}

		In this study, the same flow field boundary conditions as those used in K$\acute{a}$m$\acute{a}$n's swimming were employed, but the D-cylinder was removed to investigate the free-stream swimming of the fish. Under these conditions, the agent fish tends to maintain its swimming position in the flow field by increasing its waving frequency. The work performed by the agent fish when it is neither moving backward nor forward in the flow field (i.e., fish work) was compared with that of the K$\acute{a}$m$\acute{a}$n gait at the same upstream flow velocity. Figures \ref{fig_18_work}(a), (b) and (c) present the transient work time history of the fish in both the K$\acute{a}$m$\acute{a}$n vortex and free-swimming conditions for three different upstream speeds, respectively, with only the first 10T being shown. The transient work and total work of the agent fish under the free-swimming conditions are found to be higher. In the free-stream swimming conditions, the agent fish is unable to extract energy from the vortex in the flow field for energy-efficient swimming and must rely on actively increasing its own waving frequency to generate greater power and maintain its position in the flow field. The results of the study reveal that during the swimming process, in working condition 1, the total fish work in the K$\acute{a}$m$\acute{a}$n gait is 1526.73, while in the free swim, it is 7577.94, representing a 496.4\% increase in total output compared to the K$\acute{a}$m$\acute{a}$n swimming condition. Similarly, in working condition 2, the total fish work in the K$\acute{a}$m$\acute{a}$n gait is 2145.74, and in the free swim, it is 14682.95, representing a 684.3\% increase in total output compared to the K$\acute{a}$m$\acute{a}$n swimming condition. In working condition 3, the total fish work in the K$\acute{a}$m$\acute{a}$n gait is 3057.03, and in the free swim, it is 51520.26, representing a 1685.3\% increase in total output compared to the K$\acute{a}$m$\acute{a}$n swimming condition. Across the three working conditions, the agent fish demonstrates energy savings of 396.4\%, 584.3\%, and 1685.3\% using the K$\acute{a}$m$\acute{a}$n gait, respectively. These results demonstrate that the agent fish can significantly save energy when using the K$\acute{a}$m$\acute{a}$n vortex, by exploiting the high fluid shear stress region of the vortex in the K$\acute{a}$m$\acute{a}$n vortex street\cite{tongStudyEnergyExtraction2021}.
		
		\section{Conclusions}
		\label{sec:conclusions}
		
		\paragraph{}
		In this study, we propose a novel numerical agent fish simulation platform incorporating a lateral- line machine and macro-action system, which allows for the simulation of different flow fields through the lateral line function and the implementation of macro-actions to maintain the agent fish's position in various turbulent flows. To accomplish this, we first simulate the interaction between the agent fish's body and the flow field using an immersed boundary-Lattice Bolzmann method, based on an iterative force correction scheme. This method enables the generation of a large amount of training data for machine learning in a relatively short period of time, while maintaining computational accuracy. However, a traditional reinforcement learning approach in fish simulation can lead to the failure of a well-trained swimming policy when transferred to an unfamiliar environment, which is a fundamental limitation of reinforcement learning. To address this issue, we have designed a knowledge transplantation system that enables the transfer of well-trained knowledge through the lateral line mechanism and macro-action space, thereby avoiding the waste of knowledge. We have employed a soft actor-critic deep reinforcement learning algorithm based on a maximum entropy objective and a stochastic policy, coupled with the lateral-line machine and macro-action system, as well as a co-designed state and reward function, to simulate the swimming behavior of real fish. The numerical simulation platform was initially tested using fish predatory swimming, and then. the well-trained knowledge was subsequently successfully transferred to three different turbulent flow fields. In the K$\acute{a}$m$\acute{a}$n gait test, the swimming policy of the fish was trained at a specific upstream speed and then applied in different flow fields. The fish activated its lateral line mechanism, enabling it to adapt to movement in various turbulent flow environments by using the K$\acute{a}$m$\acute{a}$n gait to match the vortex shedding frequency and implement complex macro-actions to maintain its position in the flow field.However, there is still room for improvement. In order to enable the fish to recognize non-constant upstream flow and achieve real-time perception of the flow field environment, it is necessary to further establish a lateral-line perception mechanism that is insensitive to the flow field's time and coordinates of the flow grid points with respect to the fish's body. With the establishment of such a mechanism, the fish will possess stronger adaptability to non-constant upstream flow processes with variable flow velocity in real-time, and the ability to adapt to unsteady flow processes with drastic changes in flow velocity. It seems that the focus of future research should be on realizing a lateral- line machine for time-sequence information, online and local flow field information perception, with the aid of more advanced artificial intelligence algorithms.

		\section*{Acknowledgments}
		This work was supported by the Program of national natural science Fundation for young scholars (52109150) and by the Science and Technology Research Program of Chongqing Municipal Education Commission(Grant no. KJQN201900748).

		\bibliographystyle{unsrt}
		\bibliography{reference}

	\end{document}